\documentclass{xjkas}

\def\year{2014} 
\def\month{October} 
\def\volume{00} 
\def\beginpage{1} 
\def\endpage{19} 
\def\received{June 11, 2014} 
\def\accepted{October 18, 2014} 
\date{Received \received; accepted \accepted}

\usepackage{flushend} 

\usepackage[breaklinks,colorlinks,citecolor=blue,linkcolor=blue,menucolor=blue,urlcolor=blue]{hyperref}

\def\aa{{\langle}}
\def\rr{{\rangle}}
\def\xvec{{{\bf x}}}
\def\dvec{{{\bf d}}}
\def\etal{et al.}
\def\deg{{^{\circ}}}

\newcommand\name[1]{{\small\sc #1}}

\title{
Optical Multi-Channel Intensity Interferometry -- or:\\ How To Resolve O-Stars in the Magellanic Clouds
}

\author{Sascha~Trippe, Jae-Young~Kim, Bangwon~Lee, Changsu~Choi, Junghwan~Oh,\\ Taeseok~Lee, Sung-Chul~Yoon, Myungshin~Im, Yong-Sun~Park}
\affil{Department of Physics and Astronomy, Seoul National University, 599 Gwanak-ro, Gwanak-gu, Seoul 151-742, Korea; \email{trippe@astro.snu.ac.kr}}

\def\corrauthor{
S. Trippe
}

\def\runningauthor{
Trippe et al.
}

\def\runningtitle{
Optical Multi-Channel Intensity Interferometry
}

\def\keywords{
Instrumentation: interferometers --- Techniques: interferometric
}

\def\abstracttext{
Intensity interferometry, based on the Hanbury Brown--Twiss effect, is a simple and inexpensive method for optical interferometry at microarcsecond angular resolutions; its use in astronomy was abandoned in the 1970s because of low sensitivity. Motivated by recent technical developments, we argue that the sensitivity of large modern intensity interferometers can be improved by factors up to approximately 25\,000, corresponding to 11 photometric magnitudes, compared to the pioneering Narrabri Stellar Interferometer. This is made possible by (i) using avalanche photodiodes (APD) as light detectors, (ii) distributing the light received from the source over multiple independent spectral channels, and (iii) use of arrays composed of multiple large light collectors. Our approach permits the construction of large (with baselines ranging from few kilometers to intercontinental distances) \emph{optical} interferometers at the cost of (very) long-baseline \emph{radio} interferometers. Realistic intensity interferometer designs are able to achieve limiting $R$-band magnitudes as good as $m_R\approx14$, sufficient for spatially resolved observations of main-sequence O-type stars in the Magellanic Clouds. Multi-channel intensity interferometers can address a wide variety of science cases: (i) linear radii, effective temperatures, and luminosities of stars, via direct measurements of stellar angular sizes; (ii) mass--radius relationships of compact stellar remnants, via direct measurements of the angular sizes of white dwarfs; (iii) stellar rotation, via observations of rotation flattening and surface gravity darkening; (iv) stellar convection and the interaction of stellar photospheres and magnetic fields, via observations of dark and bright starspots; (v) the structure and evolution of multiple stars, via mapping of the companion stars and of accretion flows in interacting binaries; (vi) direct measurements of interstellar distances, derived from angular diameters of stars or via the interferometric Baade--Wesselink method; (vii) the physics of gas accretion onto supermassive black holes, via resolved observations of the central engines of luminous active galactic nuclei; and (viii) calibration of \emph{amplitude} interferometers by providing a sample of calibrator stars.
}

\begin{document}
\jkashead 



\section{Introduction \label{sect_intro}}

\begin{quote}
\small
Non est ad astra mollis e terris via.\\
There is no easy way from the Earth to the stars.

\hfill ---  Seneca the Younger (c. 54), \emph{Hercules Furens} 
\end{quote}
The demand for ever higher angular resolution in astronomy has driven the development of interferometric techniques since the mid of the 19th century (e.g., \citealt{fizeau1868,stephan1874,michelson1891}). Almost all astronomical interferometers ever constructed are \emph{amplitude} interferometers, based on the coherent superposition of electromagnetic \emph{waves}. A second, less well known, approach is provided by \emph{intensity} interferometry, based on correlations of fluctuations in the \emph{intensities} of radiation: the \emph{Hanbury Brown--Twiss effect}.

The history of \emph{Hanbury Brown--Twiss intensity interferometry} (HBTII) may well be regarded as a ``perfect science story'':

\noindent {$\bullet$} from the discovery of a phenomenon: correlated intensity fluctuations in the light received from an astronomical source of radiation \citep{hanbury1952,hanbury1954};\footnote{We note that {\scriptsize\sc Hanbury Brown}, without hyphenation, is the correct writing of the scholar's family name.}

\noindent {$\bullet$} to the development of the corresponding theory: photon-photon correlation in coherent beams of radiation \citep{hanbury1957a};

\noindent {$\bullet$} to the experimental validation of the theory \citep{hanbury1957b,twiss1959}; 

\noindent {$\bullet$} and, eventually, toward an astronomical application: optical stellar interferometry \citep{hanbury1958a,hanbury1958b}.

The discoveries of the 1950s evolved into the construction of a dedicated observatory, the Narrabri Stellar Intensity Interferometer (NSII) in Narrabri, Australia \citep{hanbury1964,hanbury1967a}. The instrument was composed of two movable 6.7-meter diameter light collectors, spanning a maximum baseline of $b=188$ meters length. As the instrument worked at a wavelength $\lambda=440$~nm, its best angular resolution was $\theta\approx1.2\lambda/b\approx0.6$ milliseconds of arc (mas). This resolving power was used for measuring directly the angular diameters of a sample of 32 bright stars \citep{hanbury1967b,hanbury1968,hanbury1974a} -- the NSII was the first astronomical instrument ever able to do so. However, despite the large collecting areas of $\approx$30~m$^2$ for each light collector, photon statistics limited the NSII to targets with apparent $B$ band magnitudes $m_B<2.5$. Accordingly, the instrument was shut down in 1972, after having completed its survey of stellar diameters. A successor was proposed around this time but not built eventually; this decision was based on the -- incorrect -- assumption that amplitude interferometry would be available for astronomy soon. A detailed review of intensity interferometry and the NSII is provided by \citet{hanbury1974b}. Today, stellar intensity interferometry is standard content of optics textbooks (cf., e.g., \citealt{fowles1975,goodman1985,mandel1995,born1999,loudon2000,kitchin2009})

In this article, we review and discuss the theory, concepts, and limitations of optical intensity interferometry. Building upon recent technical developments we argue that the sensitivity of an NSII-type interferometer can be improved by a factor of approximately 100, corresponding to 5 photometric magnitudes. This is possible by using (i) avalanche photodiodes (APD) as light detectors, and (ii) distributing the light received from the source over multiple spectral channels. When deploying interferometer arrays with multiple large light collectors equivalent to modern long-baseline radio interferometers, it is possible to achieve limiting $R$-band magnitudes up to $m_R\approx14$, corresponding to an improvement in sensitivity by a factor $\approx$25\,000 relative to the Narrabri interferometer. Such multi-channel intensity interferometers would provide new opportunities for observational astronomy in a variety of science cases ranging from stellar physics to the physics of supermassive black holes.

\section{Amplitude Interferometry \label{sect_ampli-interf}}

\subsection{Interference and Coherence \label{ssect_coherence}}

The concept of amplitude interferometry is based on the superposition of electromagnetic waves. (The following discussion is based on: \citealt{born1999}; \citealt{loudon2000}; \citealt{labeyrie2006}; \citealt{glindemann2011}.) If this superposition is coherent, the waves interfere; the interference pattern can be analyzed to obtain the properties of the source of radiation. The degree of coherence of two electric field waves $E$ observed at times $t_i$ and positions $\xvec_i$ (located in a plane perpendicular to the direction of light propagation), with $i=1,2$, is quantified by the complex \emph{first-order coherence function}
\begin{equation}
\gamma(\xvec_1,t_1,\xvec_2,t_2) = \frac{\aa E^*(\xvec_1,t_1)E(\xvec_2,t_2)\rr }{\sqrt{\aa |E(\xvec_1,t_1)|^2\rr\aa |E(\xvec_2,t_2)|^2\rr }}
\label{eq_g1}
\end{equation}
where $^*$ marks a complex conjugate, and $|...|$ and $\aa...\rr$ denote the absolute value and the time average of the enclosed expressions, respectively. By construction, $|\gamma|\in[0,1]$, where a value of 1 denotes coherent light, a value of 0 denotes incoherent light, and intermediate values denote partial coherence. Being a complex function, $\gamma$ can be expressed as $\gamma=|\gamma|\exp(i\phi)$, with $\phi$ denoting the phase. As $\gamma$ is a function of time and position, coherence is referred to as temporal coherence or spatial coherence depending on context. For convenience, $\gamma$ is usually expressed as function of differential positions and times, i.e., $\gamma(\xvec_1,t_1,\xvec_2,t_2)\rightarrow\gamma(\dvec,\tau)$, with $\dvec=\xvec_1-\xvec_2$ and $\tau=t_1-t_2$.

The necessity of \emph{temporal} coherence demands that the time delay $\tau$ between two light beams has to be much smaller than their \emph{coherence time} $\tau_c$ -- i.e., $\tau\ll\tau_c$ -- to warrant $|\gamma|\approx1$. For $\tau\gg\tau_c$, $|\gamma|=0$, and interference is not possible. For light with an optical bandwidth $\Delta\nu$, with $\nu$ being the frequency of radiation, the coherence time is given by
\begin{equation}
\tau_c \approx \frac{1}{\Delta\nu} ~.
\label{eq_tauc}
\end{equation}
The coherence time can be expressed in terms of the \emph{coherence length} $w_c=c\tau_c$, with $c$ denoting the speed of light. Evidently, any path difference $w$ between the light beams has to fulfill $w\ll w_c$ in order to preserve coherence.

The conditions for \emph{spatial} coherence are expressed by the \emph{van Cittert--Zernicke theorem}. The theorem states that the intensity distribution of an incoherent, quasi-monochromatic source and the coherence function are a Fourier transform pair. For a two-dimensional astronomical source with angular coordinates $\theta_1, \theta_2$ and sky intensity distribution $I(\theta_1,\theta_2)$, this may be written like
\begin{equation}
\gamma(\dvec) \stackrel{\mathcal{F}}{\rightleftharpoons} I(\theta_1,\theta_2)
\label{eq_cittert}
\end{equation}
where $\mathcal{F}$ denotes a Fourier transform. Accordingly, the image of a source can be derived by measuring the complex coherence function and applying a Fourier transform to it. An example important in astronomy is the case of a circular source with angular diameter $\theta$, for which
\begin{equation}
|\gamma(d)| = \left|\frac{2J_1(\zeta)}{\zeta}\right|
\label{eq_disk}
\end{equation}
where $J_1$ is the Bessel function of first order and first kind, $\zeta=\pi d\theta/\lambda$, and $d=|\dvec|$. This $|\gamma(d)|$ reaches zero value for the first time at $\zeta=3.83$, corresponding to $\theta=1.22\,\lambda/d$, leading to the well-known resolution criterion of Rayleigh. The distribution of intensity, $|\gamma(d)|^2$, then corresponds to the well-known Airy intensity profile.

The general coherence function $\gamma(\dvec,\tau)$ is conveniently expressed as function of the $uv$ coordinates ${\bf u}=(u,v,w)$. Any astronomical interferometer is an array of two or more telescopes. For each pair of telescopes, their spatial separation is given by a physical \emph{baseline} vector ${\bf b}=(b_x,b_y,b_z)$ which is defined relative to the local Earth tangential plane. By convention, $b_x$ is the telescope distance in north-south direction, $b_y$ is the distance in east-west direction, and $b_z$ is the distance in vertical direction, i.e., a height difference. By construction, $d\leq b$ with $b=|{\bf b}|$. The vector {\bf u}, in units of observation wavelength\footnote{In this convention, ${\bf u}$ is unit free, albeit it may be expressed in inverse angular units (radians$^{-1}$, mas$^{-1}$, etc.). Some authors omit the division by $\lambda$; in this case, ${\bf u}$ has the unit of a length.} $\lambda$, is derived by projection of ${\bf b}$ onto the plane of the sky at the position of the target. For the projection formula, see, e.g., \citet{segransan2007}.

For one given time and wavelength, each pair of telescopes provides a measurement of $\gamma$ in the plane spanned by $u$ and $v$, $\gamma(u,v)$. The van Cittert--Zernicke theorem relates $\gamma(u,v)$ and $I(\theta_1,\theta_2)$ like
\begin{equation}
\gamma(u,v) \stackrel{\mathcal{F}}{\rightleftharpoons} I(\theta_1,\theta_2) ~ .
\label{eq_uvtheta}
\end{equation}
Accordingly, we may define a \emph{uv radius} $\rho$ via $\rho^2=u^2+v^2=(d/\lambda)^2$. In this notation, the angular resolution of a baseline is then $\theta\approx1/\rho$. The coordinate $w$ corresponds to the path difference, and thus the delay, between the light rays arriving at the two telescopes. This delay needs to be controlled carefully to ensure $w\ll w_c$. By symmetry (because the selection of the origin of ${\bf b}$ is arbitrary), each measurement actually provides two values, $\gamma(u,v)$ and $\gamma(-u,-v)=\gamma^*(u,v)$.

\subsection{Radio Astronomy \label{ssect_radio}}

Radio interferometry \citep{thompson2004,wilson2010} is based on preserving the full wave (amplitude and phase) information of the infalling radiation while recording it. This is achieved by \emph{heterodyne receivers} that shift the signal frequency to lower values by \emph{mixing} the astronomical signal with a stable reference wave provided by a \emph{local oscillator}. The output electric voltage can be transmitted and processed almost arbitrarily by use of dedicated electronics. Especially, it is straightforward to correlate the signals from different antennas and to calculate $\gamma({\bf u})$ according to Equation (\ref{eq_g1}).

The technical simplicity of radio interferometry made it a standard technique in observational astronomy. Traditionally, radio interferometers are classified according to their physical extensions: \emph{long-baseline interferometry} (LBI) denotes interferometry with arrays composed of several, physically separate telescopes. Signals from different antennas are usually correlated and processed in real time. \emph{Very long baseline interferometry} (VLBI) refers to arrays that are too extended -- potentially extending across the entire Earth -- to combine the signals from individual telescopes in real time. In VLBI arrays, the signal of any antenna is stored locally by powerful (magnetic tape or hard disk) recorders, together with an accurate time reference signal usually provided by an atomic clock. The amplitudes from individual antennas are then correlated off-line.

\subsection{Optical Astronomy \label{ssect_optical}}

At wavelengths shorter than a few micrometers, fundamental quantum noise limits prevent the use of heterodyne receiving techniques.\footnote{At least in astronomy. In laboratory optics, where light sources of almost arbitrary intensities are available, optical heterodyne detection is a standard experimental technique (e.g., \citealt{bachor2004}). At $\lambda\approx11.2\,\mu$m, heterodyne detection has been employed by the Berkeley Infrared Spatial Interferometer (ISI) which uses CO$_2$ lasers as local oscillators \citep{hale2000}.} In order to achieve interference, the light from two (or more) apertures has to be combined \emph{directly} by superposition \citep{labeyrie2006,glindemann2011}. Physical information is obtained by recording and analysis of the resulting interference pattern. The fringe contrast, or \emph{visibility}, obeys $V\propto|\gamma({\bf u})|$. The missing information on the phase $\phi$ is obtained from the intensity at the center position ($\rho=0$) of the fringe pattern. This results in a \emph{complex} visibility ${\mathcal V} \equiv V\exp(i\phi)\propto\gamma({\bf u})$.

The direct combination of light rays imposes severe constraints on the stability and mechanical tolerances of an optical interferometer. As stated by Equation (\ref{eq_tauc}), a wide optical bandpass leads to very short coherence times. Assuming observations at $\lambda=500$\,nm through a narrow band filter with bandpass $\Delta\lambda=10$\,nm, we find $\Delta\nu=1.2\times10^{13}$\,Hz. From this one finds a correlation time of $\tau_c\approx10^{-13}$\,s, corresponding to a correlation length $w_c\approx30\,\mu$m. As interference requires $w\ll w_c$, we need to control the geometry of our interferometer within accuracies on the order of few micrometers. This may be compared to the case of radio interferometry with typical $\Delta\nu\approx1$\,GHz, resulting in $w_c\approx0.3$\,m.

A further complication arises from atmospheric turbulence. Fluctuations of the atmosphere lead to random variations of the phase of the fringe pattern on time scales of few milliseconds. For exposure times longer than this atmospheric coherence time, the fringe pattern, and thus the visibility, is averaged out. Additionally, radiation propagating through the atmosphere remains coherent only on spatial scales on the order of 10\,cm. The use of telescope apertures substantially larger than the atmospheric coherence scale further weakens the interference pattern.

\section{Intensity Interferometry \label{sect_intensity-interf}}

\subsection{Intensity Correlations \label{ssect_correlations}}

\begin{figure}[t!]
\centering
\includegraphics[height=82mm,angle=-90]{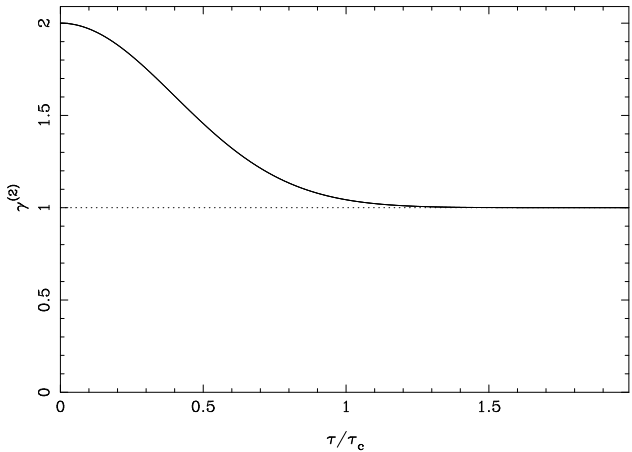}
\caption{The second-order coherence function $\gamma^{(2)}$, as function of time delay $\tau$ in units of coherence time $\tau_c$. The continuous curve indicates the case of chaotic light with Gaussian frequency profile. The dotted line with a constant value of unity corresponds to the behavior of monochromatic coherent light (laser, maser).\label{fig_g2}}
\end{figure}

Optical intensity interferometry exploits correlated fluctuations in the intensity of the radiation received from a source. Else than for optical amplitude interferometry, the radiation arriving at any antenna (telescope) is converted into an electronic signal by a detector immediately upon reception at the antenna. Only the electronic signals -- not the light rays -- are transmitted to a correlator and combined; accordingly, an intensity interferometer may be regarded as a cross of optical and radio interferometers, with an optical \emph{frontend} -- light collectors, optics, photodetectors -- and a radio \emph{backend} -- signal transmission, electronics, correlators. The electronic signal from each antenna carries information on the intensity $I(t)$ of the recorded light.\footnote{Strictly speaking, we would have to distinguish between the intensity of the infalling \emph{light} and the intensity of the recorded electronic \emph{signal} from now on. As these two intensities are equivalent throughout our analysis, we omit this distinction for simplicity.}

\begin{figure*}[!t]
\centering
\includegraphics[width=83mm]{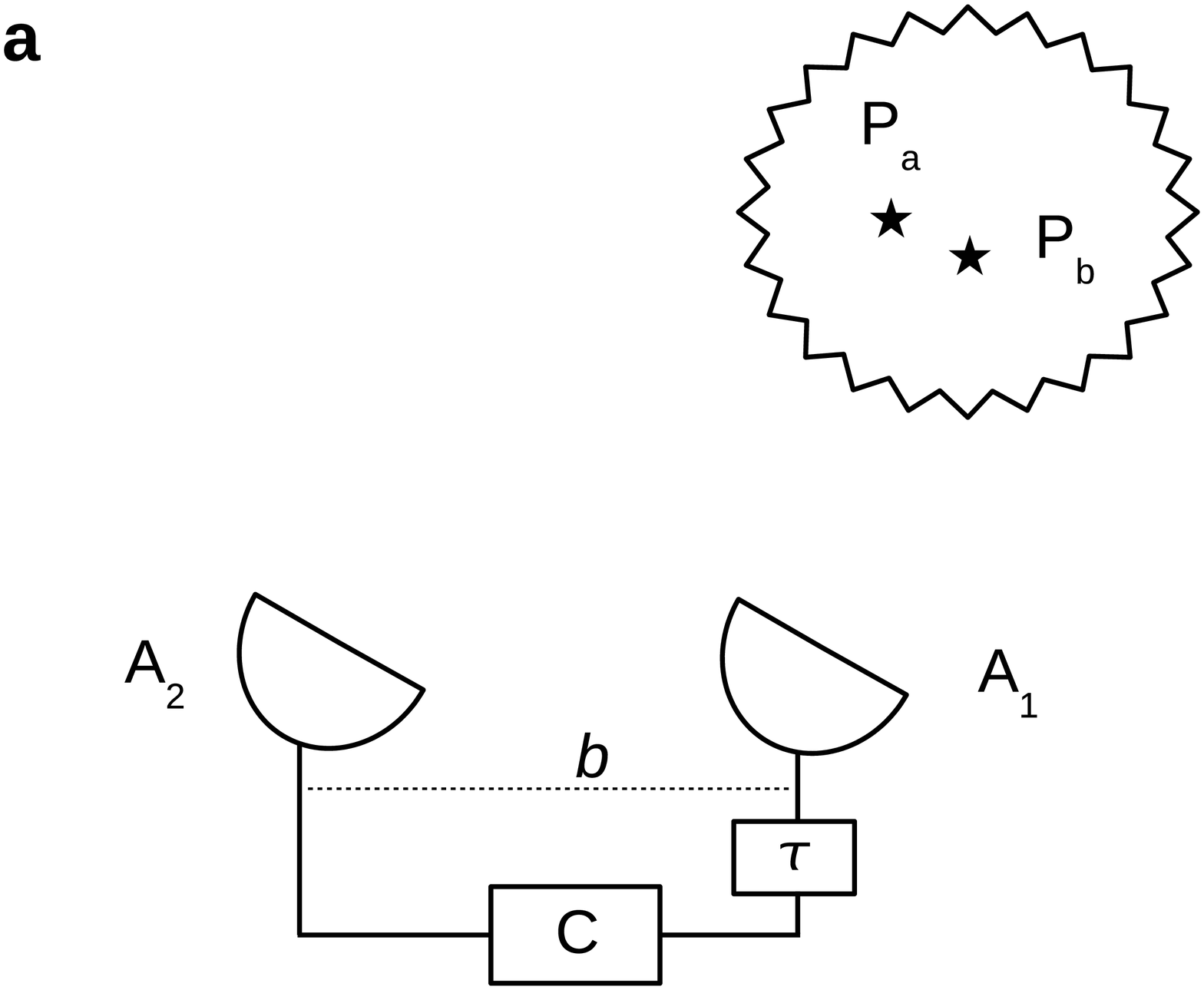}
\vline 
\hspace{4mm}
\includegraphics[width=83mm]{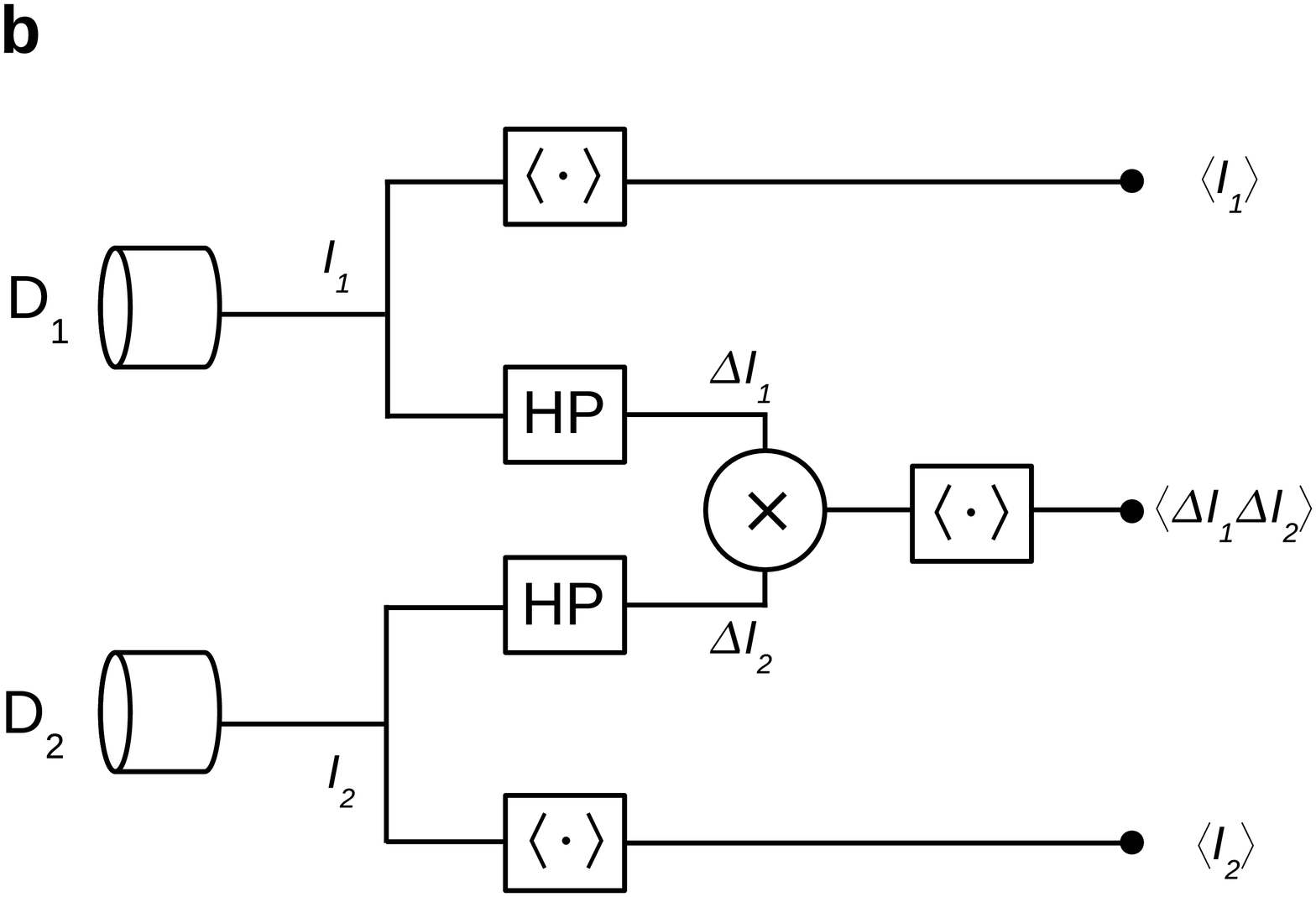}
\vspace{2mm}
\caption{Basic concepts of intensity interferometry. {\bf a}: Observation layout. Two antennas A$_{1,2}$ located at a distance $b$ observe a target. Each antenna observes an intensity given by a superposition of electric waves emitted by pairs of point sources P$_{a,b}$. At each antenna, the intensity is recorded and converted into an electronic signal. The signals are combined by a correlator C. Optical path differences are compensated by an electronic delay $\tau$ in one of the interferometer arms. {\bf b}: Signal processing scheme. Two detectors D$_{1,2}$ located in antennas A$_{1,2}$ record voltages with intensities $I_{1,2}(t)$. Each signal is passed through a high-pass filter HP that suppresses the constant direct current component, leaving the fluctuating components $\Delta I_{1,2}(t)$. The fluctuation signals are multiplied (device $\times$) and averaged in time (device $\aa\cdot\rr$), resulting in the averaged product $\aa\Delta I_1\Delta I_2\rr$ at the output. Separately, the time averaged intensities $\aa I_{1,2}\rr$ are computed and stored.\label{fig_ii}}
\end{figure*}

Intensity interferometry exploits the fact that intensities received at different antennas show correlated fluctuations. This can be understood \citep{hanbury1957a,hanbury1974b} by regarding the electromagnetic waves emitted by a pair of point sources P$_{a,b}$ with amplitudes $E_{a,b}$, angular frequencies $\omega_{a,b}$, and phases $\phi_{a,b}$ (see also Figure \ref{fig_ii}a). The intensities recorded at two antennas A$_{1,2}$ are then
\begin{eqnarray}
I_1 &=& \big[ E_a\sin(\omega_at+\phi_a) + E_b\sin(\omega_bt+\phi_b) \big]^2 \nonumber \\
I_2 &=& \big[ E_a\sin(\omega_a(t+\tau_a)+\phi_a) \nonumber \\
    & & +\,  E_b\sin(\omega_b(t+\tau_b)+\phi_b) \big]^2
\label{eq_intwave}
\end{eqnarray}
where $\tau_{a,b}$ are delays due to optical path differences between the two antennas. Each of the two expressions results in a sum of a constant direct current component plus oscillating terms that correspond to harmonics, sums, and differences of the two frequencies. By application of appropriate high-pass and low-pass filters, it is possible to suppress all terms except the one corresponding to the difference of $\omega_a$ and $\omega_b$, eventually leaving us with two beat terms
\begin{eqnarray}
\Delta I_1 &=& E_aE_b\cos\big[ (\omega_a-\omega_b)t + (\phi_a-\phi_b) \big] \nonumber \\
\Delta I_2 &=& E_aE_b\cos \big[ (\omega_a-\omega_b)t + (\phi_a-\phi_b) \nonumber \\
           & & +\, \omega_a\tau_a - \omega_b\tau_b \big]
\label{eq_intfluct}
\end{eqnarray}
Evidently, $\Delta I_1$ and $\Delta I_2$ are identical except of a phase term $\omega_a\tau_a - \omega_b\tau_b$. Correlating, i.e., multiplying the two intensities, and assuming $\omega_a\approx\omega_b\equiv\omega$, results in
\begin{equation}
\Delta I_1\Delta I_2 = E_a^2E_b^2\cos\left[\omega(\tau_a-\tau_b)\right]
\label{eq_intcorr}
\end{equation}
or, assuming the points $P_{a,b}$ have angular distance $\theta$,
\begin{equation}
\Delta I_1\Delta I_2 = E_a^2E_b^2\cos\left[2\pi b\theta/\lambda\right] ~ .
\label{eq_intcorrtheta}
\end{equation}
Integrating Equation (\ref{eq_intcorrtheta}) over all possible pairs of points P$_{a,b}$, all frequencies $\omega_{a,b}$, and all frequency differences $\omega_a-\omega_b$ for a given target source leads to the conclusion \citep{hanbury1957a} that the product $\Delta I_1\Delta I_2$ is proportional to $|\gamma|^2$. Accordingly, the correlation of fluctuations of intensities observed at antennas A$_{1,2}$ provides information on the source structure.

\subsection{Second-Order Coherence \label{ssect_2nd-order-coherence}}

In analogy to Equation (\ref{eq_g1}) for electromagnetic \emph{waves}, the degree of correlation among \emph{intensities} $I$ is quantified \citep{loudon2000,labeyrie2006} by the \emph{second-order coherence function}
\begin{equation}
\gamma^{(2)}(\dvec,\tau) = \frac{\aa I(\xvec,t)I(\xvec+\dvec,t+\tau) \rr}{\aa I(\xvec,t)\rr\aa I(\xvec+\dvec,t+\tau)\rr}
\label{eq_g2}
\end{equation}
where the superscript $^{(m)}$ denotes coherence of $m$-th order ($m>1$); $\dvec$ and $\tau$ are the parameters already used in Section \ref{ssect_coherence}. The \emph{temporal} second-order coherence, $\gamma^{(2)}\equiv\gamma^{(2)}(\tau)$ (i.e., $\dvec=0$), is a function of the photon statistics of the radiation to be analyzed \citep{loudon2000,bachor2004}. For \emph{monochromatic coherent light} with Poissonian photon statistics, as provided by lasers or masers, $\gamma^{(2)}(\tau)=1$ for all $\tau$. This is distinct from the case of \emph{chaotic} light, referring to radiation composed of electric waves with random amplitudes that follow a Gaussian distribution; this includes especially thermal black-body radiation. Chaotic light obeys \emph{super-Poissonian} statistics, commonly referred to as \emph{photon bunching}, resulting in characteristic $\gamma^{(2)}(\tau)$ profiles. In general, $\gamma^{(2)}(0)=2$ and $\gamma^{(2)}(\tau\gg\tau_c)=1$, with $\tau_c$ being the coherence time according to Equation (\ref{eq_tauc}). The exact shape of the $\gamma^{(2)}(\tau)$ curve depends on the frequency profile of the light (see, e.g., Chapter 3.7 of \citealt{loudon2000} for details). In Figure \ref{fig_g2}, we show the $\gamma^{(2)}(\tau)$ profile of chaotic light with Gaussian frequency profile (\emph{Gaussian--Gaussian light}) alongside the trivial case of coherent light.\footnote{In addition, there is the case of \emph{sub-Poissonian} radiation with $\gamma^{(2)}(0)<1$, also referred to as \emph{photon anti-bunching} or \emph{non-classical light}. This type of light is irrelevant in astronomy.}

For the case of chaotic light, a generalization of the analysis introduced in Section \ref{ssect_correlations} finds a general relation between the \emph{first-order} coherence function $\gamma(\dvec,\tau)$ and the \emph{second-order} coherence function $\gamma^{(2)}(\dvec,\tau)$ given by
\begin{equation}
\gamma^{(2)}(\dvec,\tau) = 1 + |\gamma(\dvec,\tau)|^2
\label{eq_g1-g2}
\end{equation}
\citep{loudon2000,labeyrie2006}. Accordingly, an analysis of the second-order coherence function provides the squared modulus of the first-order coherence function. Notably, this statement holds for both, temporal and spatial coherence. This means that an intensity interferometer provides information on the spatial structure of the target source similar to an amplitude interferometer. The most important difference is the loss of phase information as only the modulus of $\gamma$ is preserved. As a consequence, it is not possible to reconstruct the complete source image by Fourier transformation of the $uv$ plane data, as this requires the complex function $\gamma(\bf u)$.

In general, each pair of antennas A$_{1,2}$ of an astronomical intensity interferometer records intensities $I_{1,2}(t)$. By subtraction of the time averages of the intensities $\aa I(t)\rr$, we obtain the intensity fluctuations around the averages, $\Delta I_{1,2}(t)=I_{1,2}(t)-\aa I_{1,2}(t)\rr$. Here time averages are taken over periods much longer than the resolving time $\tau_d$ of the light detectors, and shorter than the time scales of any variability of the source brightness. By comparison of Equation (\ref{eq_g2}) and Equation (\ref{eq_g1-g2}), one finds the convenient relation
\begin{equation}
|\gamma(\dvec,\tau)|^2 = \frac{\aa\Delta I_1\Delta I_2\rr}{\aa I_1\rr\aa I_2\rr}
\label{eq_g1sq}
\end{equation}
which provides $|\gamma(\dvec,\tau)|^2$ from correlation of the intensity \emph{fluctuations}. In Figure \ref{fig_ii}b, we present the corresponding signal processing scheme of an intensity interferometer (see also, e.g., Chapter 6.3.1 of \citealt{goodman1985}). Two detectors D$_{1,2}$ record voltages with intensities $I_{1,2}$. Each intensity is averaged in time, providing $\aa I_{1,2}\rr$ at the corresponding outputs. The intensity fluctuations $\Delta I_{1,2}$ are obtained by passing the signals through electronic high-pass filters that suppress (subtract) the direct current components $\aa I_{1,2}\rr$. The intensity fluctuations are multiplied and the product is averaged in time, eventually providing $\aa\Delta I_1\Delta I_2\rr$. From the various output parameters it is straightforward to compute the \emph{normalized correlation}
\begin{equation}
c(u,v) = \frac{\aa\Delta I_1\Delta I_2\rr}{\aa I_1\rr\aa I_2\rr}
\label{eq_corr}
\end{equation}
Obviously, Equation (\ref{eq_g1sq}) and Equation (\ref{eq_corr}) are equivalent; our correlation analysis simply provides the squared modulus of the first-order coherence function as function of $uv$ plane location (the delay $w$ will be discussed in Section \ref{ssect_coherencetimes}).

At this point, we encounter a difficulty which arises from the properties of photo detectors. As already discussed in Section \ref{ssect_optical}, we have to deal with coherence times $\tau_c\lesssim10^{-13}$\,s in optical interferometry. The shortest resolving times $\tau_d$ are provided by photomultiplier tubes (PMT) and avalanche photodiodes (APD) which are known from various applications in astronomy \citep{renker2007,kitchin2009}; modern, commercially available, versions of these detectors\footnote{See, e.g., the product catalog of {\sc Hamamatsu Photonics}, Hamamatsu; {\tt http://jp.hamamatsu.com/en/product\_info}} provide $\tau_d\gtrsim5\times10^{-10}$\,s. Comparison with Figure \ref{fig_g2} shows that for $\tau\geq\tau_d\gg\tau_c$, observed correlations $c(u,v)$ will be much smaller than unity, and only slightly larger than zero, for all $u, v$. Accordingly, a more practical parameter is the \emph{normalized correlation factor}
\begin{equation}
\Gamma(u,v) = \frac{c(u,v)}{c(0,0)}
\label{eq_gamma}
\end{equation}
which expresses the correlation $c(u,v)$ in units of the correlation obtained at the origin of the $uv$ plane.

\begin{figure}[t!]
\centering
\includegraphics[height=82mm,angle=-90]{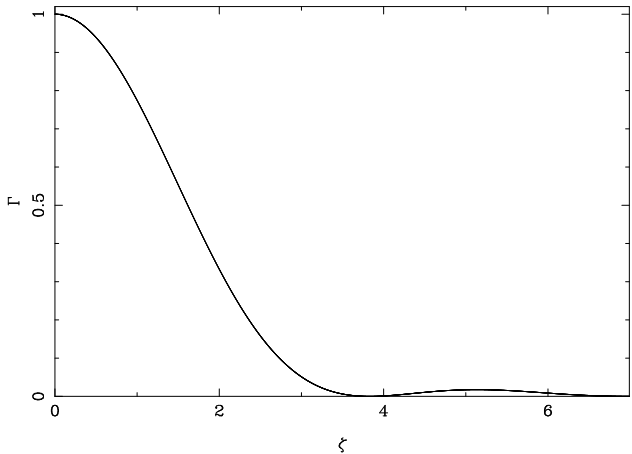}
\vspace{2mm}
\caption{Correlation factor $\Gamma$ as function of $\zeta=\pi\theta\rho=\pi\theta d/\lambda$, if the target source is a uniform circular disk. The first zero point, where the source is resolved entirely, is located at $\zeta=3.83$, corresponding to $\theta=1.22\lambda/d$. For example: at $\lambda=700$\,nm, a star with angular diameter $\theta=1$\,mas is resolved entirely by a projected baseline of $d=176$\,m.\label{fig_disk}}
\end{figure}

\begin{figure*}[!t]
\centering
\includegraphics[width=84mm]{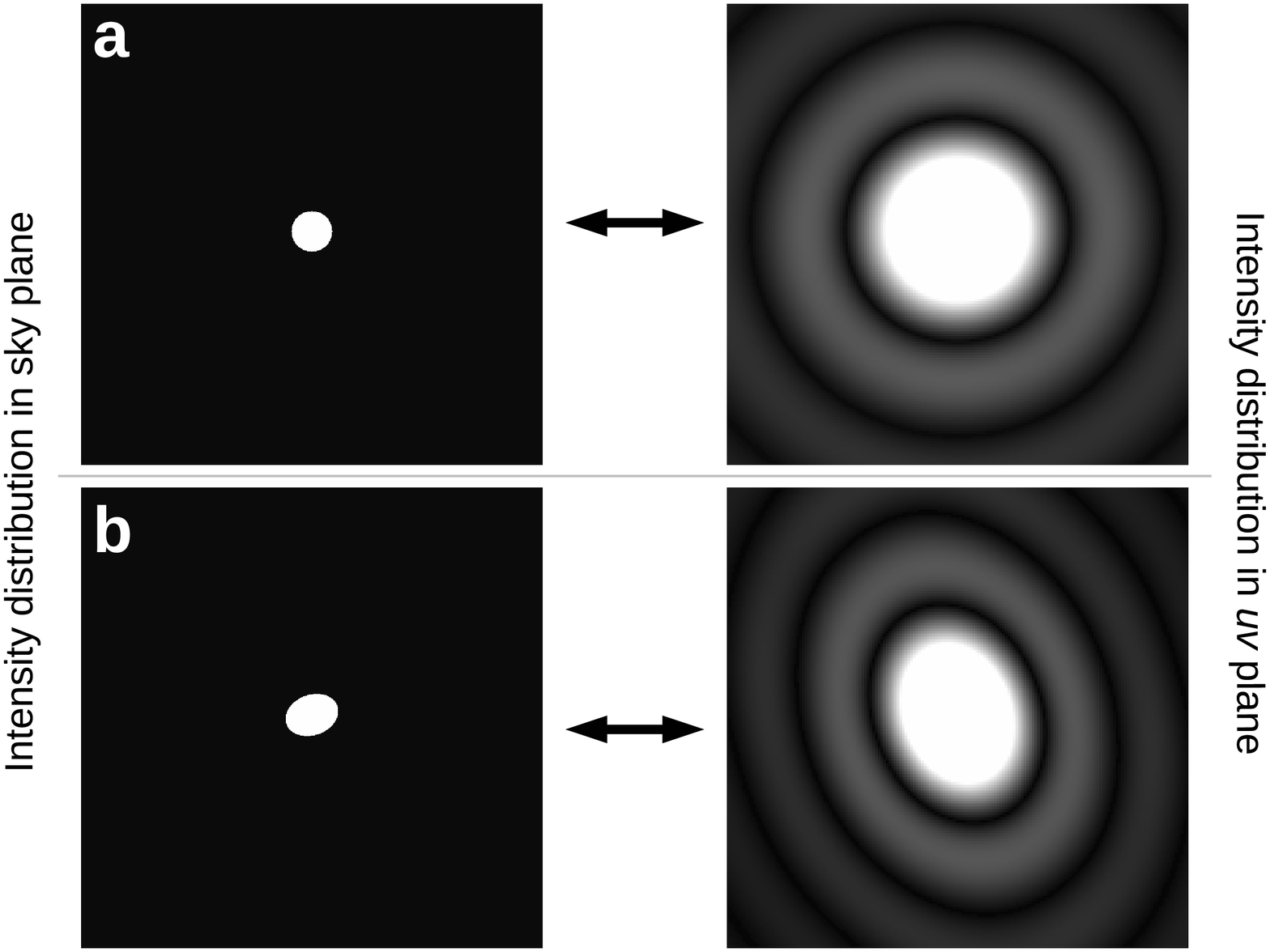}
\hspace{0mm}
\vline
\hspace{1mm}
\includegraphics[width=84mm]{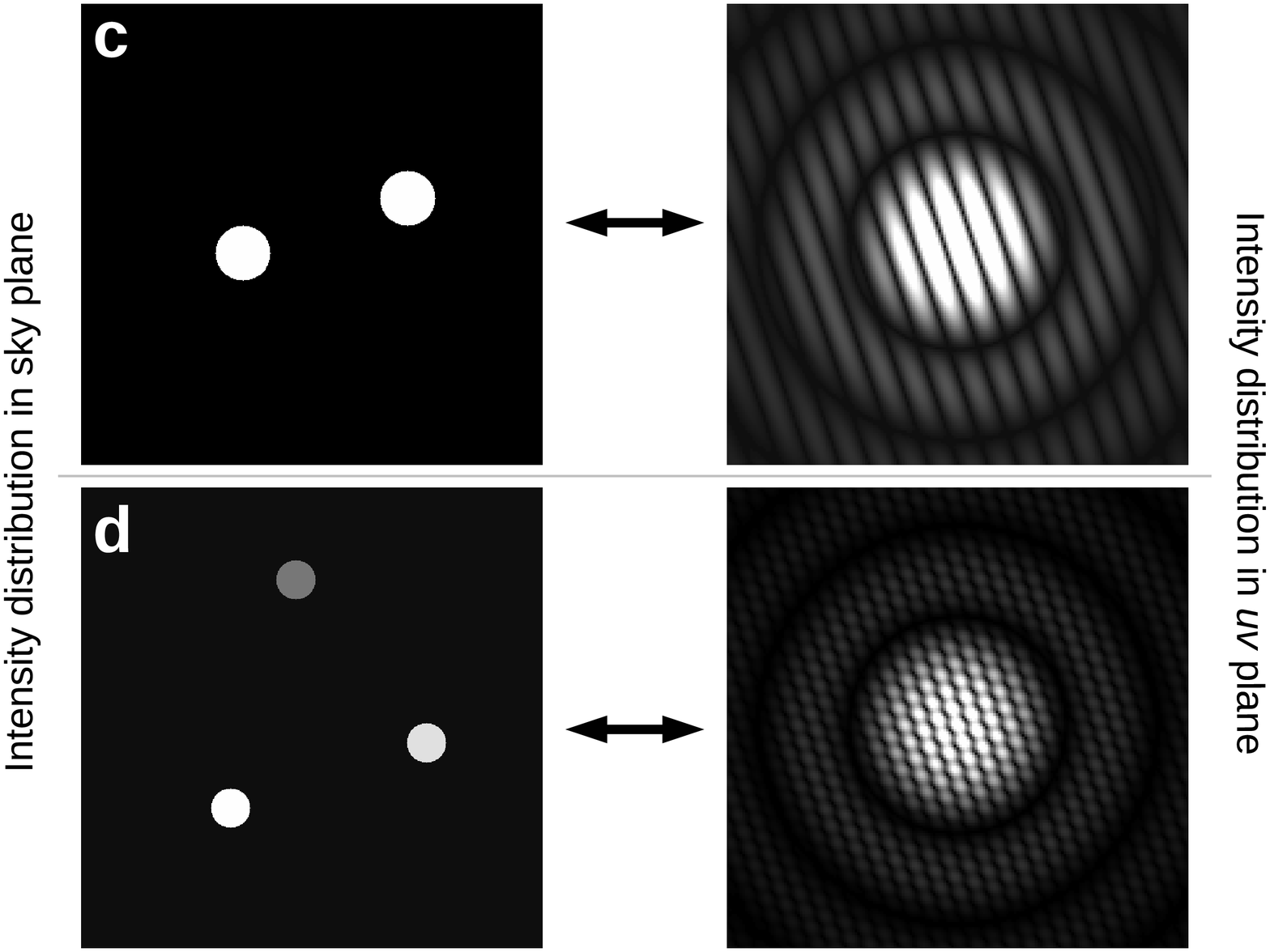}
\vspace{2mm}
\caption{Sky plane vs. \emph{uv} plane intensity distributions for various target source geometries. All scales are arbitrary. {\bf a}: A disk-like circular star; the profile shown in Figure \ref{fig_disk} corresponds to a 2-dimensional cut through the $uv$ plane distribution shown here. {\bf b}: An elliptical, rotation-flattened star, with axis ratio $4:3$. {\bf c}: A binary system of two disk-like circular stars of equal brightness and size, separated by 3.2 times the stellar diameter. {\bf d}: A triple star system with relative brightnesses (stars from bottom to top) $3:2:1$.\label{fig_multi}}
\end{figure*}

As $\Gamma(u,v)\propto|\gamma(u,v)|^2$ by construction, the normalized correlation factor encodes the spatial source structure on a convenient scale. Accordingly, the relation between $\Gamma(u,v)$ and the spatial intensity distribution of an astronomical source are given by the van Cittert--Zernicke theorem, however with the limitation that phase information is not preserved. The source structure is found by fitting an appropriate parametric model to the $\Gamma(u,v)$ data, with the model being the normalized square modulus of the Fourier transform of the theoretical source structure \citep{hanbury1958a,hanbury1958b}. The arguably most important example is the measurement of stellar diameters with the NSII \citep{hanbury1974b}: when approximating a star as a uniform circular disk, the corresponding model for $\Gamma$ is an Airy profile as obtained by taking the square of Equation (\ref{eq_disk}); the angular diameter of the stellar disk is then derived from measurements of $\Gamma(\rho)$, with $\rho=d/\lambda$ being the $uv$ radius. The first zero point of the Airy profile -- marking the point where a source is resolved entirely, according to the Rayleigh criterion -- corresponds to an angular disk diameter of $\theta=1.22\lambda/d$; for example: at $\lambda=700$\,nm, a star with angular diameter $\theta=1$\,mas is resolved entirely by a projected baseline of $d=176$\,m. For the Narrabri interferometer, $\lambda=440$\,nm, $d=188$\,m at most, and accordingly the resolution limit was $\theta=0.59$\,mas \citep{hanbury1967a}. We present the profile of $\Gamma(\rho)$ for the case of a uniform circular disk source in Figure \ref{fig_disk}. In Figure \ref{fig_multi} we show examples for the $\Gamma(u,v)$ distributions of (a) a disk-like, circular star; (b) a rotation-flattened, elliptical star; (c) a close binary system; and (d) a close triple star system, respectively.

The phase information, which is lost by correlating the intensities recorded at two antennas, may be retrieved by means of a \emph{triple correlation} \citep{gamo1963,sato1978}. This approach makes use of \emph{third-order coherence} which relates the signals from \emph{three} antennas like
\begin{equation}
\Gamma^{(3)} \equiv \frac{\aa\Delta I_1\Delta I_2\Delta I_3\rr}{\aa I_1\rr\aa I_2\rr\aa I_3\rr} ~ .
\label{eq_g3}
\end{equation}
For a given triple of antennas $(1,2,3)$, the parameter $\Gamma^{(3)}$ is related to the first-order coherence function $\gamma$ like
\begin{equation}
\Gamma^{(3)} = a|\gamma_{12}||\gamma_{13}||\gamma_{23}|\cos(\psi)
\label{eq_triple}
\end{equation}
where $a$ is a constant, $\gamma_{ij}$ is the first-order coherence function derived for the antenna pair $ij$, and $\psi$ is given by
\begin{equation}
\psi = \phi_{12} - \phi_{13} + \phi_{23}
\label{eq_psi}
\end{equation}
where $\phi_{ij}\equiv-\phi_{ji}$ denotes the phase of $\gamma_{ij}$. The relations given in Equations (\ref{eq_triple}) and (\ref{eq_psi}) are equivalent to the \emph{closure phase} relations known from amplitude interferometry \citep{jennison1958}. For an array of $M$ antennas, the number of unknown phases equals the number of baselines $N_{\Join}=M(M-1)/2$, whereas the number of independent triangles of antennas, and thus values of $\Gamma^{(3)}$, is $N_{\bigtriangleup}=(M-1)(M-2)/2$. As $N_{\Join}-N_{\bigtriangleup}=M-1$, the number of unknowns always exceeds the number of closure phases; a reliable determination of the $\phi_{ij}$ requires either large $M$ or the presence of $M-3$ redundant baselines \citep{labeyrie2006}. As noted by \citet{holmes2004,nunez2012a,nunez2012b}, the phase information can also be recovered by application of the Cauchy--Riemann equations to the $uv$ data, sufficient sampling of the $uv$ plane provided.

\subsection{Coherence Times and Tolerances \label{ssect_coherencetimes}}

The coherence time, and thus the coherence length, relevant for intensity interferometers derive from the properties of the photodetectors and the signal processing system. The \emph{maximum} frequency of an excitation that can be recorded by photodetectors is found from the Nyquist theorem:
\begin{equation}
f_{\rm max} = \frac{1}{2\tau_d}
\label{eq_fmax}
\end{equation}
with $\tau_d$ denoting the time resolution of the detector.\footnote{By convention, technical descriptions of photodetectors sometimes refer to the \emph{impulse rise time} $\tau_r$, for which $f_{\rm max}\approx0.35/\tau_r$.} From comparison to Equation (\ref{eq_intfluct}) it is evident that $f_{\rm max}$ corresponds to the highest beat frequency $(\omega_a-\omega_b)$ observable by a detector. For a realistic signal processing system (see Section \ref{ssect_correlations} and Figure \ref{fig_ii}b), only signals with frequencies above a certain \emph{minimum} frequency $f_{\rm min}$ will be processed by the system; frequencies lower than $f_{\rm min}\lesssim1$\,MHz will be suppressed by high-pass filters. Eventually, we obtain the \emph{electronic bandwidth}
\begin{equation}
\Delta f = f_{\rm max} - f_{\rm min}
\label{eq_band}
\end{equation}
which is simply the difference of maximum and minimum signal frequencies. For practical instruments, $f_{\rm max}\gg f_{\rm min}$, and, accordingly, $\Delta f\approx f_{\rm max}$.

The \emph{electronic} bandwidth $\Delta f$, rather than the \emph{optical} bandwidth $\Delta\nu$, defines the coherence time of an intensity interferometer
\begin{equation}
\tau'_c \approx \frac{1}{\Delta f}
\label{eq_taucii}
\end{equation}
and, accordingly, the coherence length $w'_c=c\tau'_c$. This expression may be compared to Equation (\ref{eq_tauc}). As noted in Section \ref{ssect_2nd-order-coherence}, realistic photodetectors have $\tau_d\gtrsim5\times10^{-10}$\,s, meaning $\Delta f\approx f_{max}\lesssim1$\,GHz. Accordingly, $\tau'_c\gtrsim10^{-9}$\,s and $w'_c\gtrsim0.3$\,m; evidently, optical intensity interferometers are \emph{highly insensitive to deviations from a perfect array geometry}. Mechanical tolerances of several centimeters are very easy to meet technically; notably, the tolerances derived here for optical intensity interferometers are very similar to the tolerances of radio amplitude interferometers (see Sections \ref{ssect_radio} and \ref{ssect_optical}) -- which represent a technology well-known since the 1950s. This may be compared to the tolerances of optical amplitude interferometers which are on the order of micrometers (see Section \ref{ssect_optical}). Likewise, optical path differences $w$ occurring during an observation need to be controlled down to levels of few centimeters (and not micrometers); this is achieved by introducing an electronic delay into one arm of the interferometer that compensates the time delay $\tau$. A comparison with Equation (\ref{eq_intcorr}) shows that the treatment of delays does not influence the results obtained, as for any realistic astronomical observation $\tau\approx\tau_a\approx\tau_b$ and $\tau_a-\tau_b\ll\tau'_c$.

The implementation of a minimum frequency $f_{\min}$ by means of electronic high-pass filters is usually the technically easiest method to separate $\Delta I$ from $\aa I\rr$, as illustrated in Figure \ref{fig_ii}b; a reasonable (though somewhat arbitrary) value is $f_{\min}\approx1$\,MHz. Using those values introduces an additional major advantage of optical intensity interferometers over optical amplitude interferometers. Optical amplitude interferometers are severely limited by atmospheric fluctuations (see also Section \ref{ssect_optical}) that occur on frequency scales $\lesssim$1\,kHz. Accordingly, any $f_{\min}\gg1$\,kHz filters out atmospheric fluctuations of phases and amplitudes -- meaning that optical intensity interferometers are \emph{insensitive to atmospheric turbulence} \citep{hanbury1974b,labeyrie2006}.

\subsection{Signal-to-Noise Limits \label{ssect_signal-to-noise}}

\subsubsection{Statistical Limit \label{sssect_statlimit}}

Intensity interferometry is based on the correlation of signals obtained by photoelectric detection of light. Accordingly, we may conduct a semi-classical statistical analysis of the photoelectron counts \citep{goodman1985,loudon2000,labeyrie2006}. Assuming the generation of $n$ photoelectrons within one detector resolution time $\tau_d\gg\tau_c$, one obtains the variance of the photoelectron count
\begin{equation}
\sigma_n^2 = n\left[ 1 + n\left(\frac{\tau_c}{\tau_d}\right)\right]
\label{eq_noise}
\end{equation}
where $\tau_c$ is the coherence time of the light. The first summand, of order $n$, corresponds to Poisson noise. The second summand, of order $n^2$, represents \emph{super-Poissonian} or \emph{wave noise}, corresponding to macroscopic fluctuations of the light intensity; these intensity fluctuations form the signal of an intensity interferometer. Accordingly, a signal-to-noise ratio ($S/N$) of unity corresponds to the case where both summands are equal, meaning
\begin{equation}
n\left(\frac{\tau_c}{\tau_d}\right) = 1 ~ .
\label{eq_sn1}
\end{equation}
To obtain a \emph{statistical} $S/N$ for an \emph{arbitrary} observing time $t$, Poisson statistics tells us that we need to scale Equation (\ref{eq_sn1}) by a factor $\sqrt{t/\tau_d}$, resulting in
\begin{equation}
\left(\frac{S}{N}\right)_{\rm s} = n\left(\frac{1}{\tau_d\,\Delta\nu}\right)\left(\frac{t}{\tau_d}\right)^{1/2}
\label{eq_snglobal}
\end{equation}
where we used $\tau_c=1/\Delta\nu$ (Equation (\ref{eq_tauc})). The number of photoelectrons, $n$, collected by a telescope within the time $\tau_d$ can be expressed as
\begin{equation}
n = \alpha\,A\,n_{\nu}\,\Delta\nu\,\eta\,\tau_d
\label{eq_nel}
\end{equation}
where $\alpha\in[0,1]$ is the quantum efficiency of the photodetectors, $A$ is the collecting area of the light collector,\footnote{For a pair of collectors with different sizes, $A$ is the geometric mean of the areas of the two collectors \citep{hanbury1974b}.} $n_{\nu}$ is the number of photons per time, per unit area, and per unit bandwidth, $\Delta\nu$ is the bandwidth, and $\eta\in[0,1]$ is the efficiency of the instrument (except detector quantum efficiency). Using also $\tau_d=1/(2\Delta f)$ (for $f_{\rm min}\ll f_{\rm max}$; Equations (\ref{eq_fmax}) and (\ref{eq_band})), we obtain
\begin{equation}
\left(\frac{S}{N}\right)_{\rm s} = \alpha\,A\,n_{\nu}\,\eta\,\left(2\,t\,\Delta f\right)^{1/2} ~ .
\label{eq_snfull}
\end{equation}

At this point, we have to consider two additional effects. Firstly, we need to multiply Equation (\ref{eq_snfull}) with the normalized correlation factor $\Gamma(u,v)$ to take into account spatial coherence. Secondly, Equation (\ref{eq_snfull}) is strictly correct only for fully polarized light; for unpolarized light -- as usual in astronomy -- the observed correlation signal is reduced by a factor of two because orthogonal polarizations are uncorrelated \citep{hanbury1974b,goodman1985}. Eventually, we obtain the signal-to-noise ratio of an intensity interferometer:
\begin{equation}
\left(\frac{S}{N}\right)_{\rm s} = \alpha\,A\,n_{\nu}\,\eta\,\Gamma(u,v)\,\left(\frac{t\,\Delta f}{2}\right)^{1/2} ~ .
\label{eq_snii}
\end{equation}
Scaling Equation (\ref{eq_snii}) to practical values, one obtains the intuitive expression
\begin{eqnarray}
\label{eq_snnsii}
\left(\frac{S}{N}\right)_{\rm s} &=& 46 \left(\frac{\alpha}{0.2}\right) \left(\frac{A}{30\,{\rm m}^2}\right)\left(\frac{\eta}{0.2}\right) \\
 &\times& \left(\frac{n_{\nu}}{9\times10^{-5}\,{\rm m}^{-2}\,{\rm s}^{-1}\,{\rm Hz}^{-1}}\right)  \nonumber \\
 &\times& \left(\frac{t}{1\,{\rm h}}\right)^{1/2}\left(\frac{\Delta f}{100\,{\rm MHz}}\right)^{1/2} \Gamma(u,v) \nonumber ~ .
\end{eqnarray}
The reference values for $\alpha$, $A$, $\eta$, and $\Delta f$ correspond to values typical for the Narrabri intensity interferometer \citep{hanbury1974b}. A photon flux of $n_{\nu}=9\times10^{-5}\,{\rm m}^{-2}\,{\rm s}^{-1}\,{\rm Hz}^{-1}$ corresponds to the photon flux from a star observed at $\lambda\approx500-800$\,nm ($V, R, I$ bands) with an apparent photometric magnitude of zero \citep{bessell1979}, attenuated by $\approx$10\% due to atmospheric extinction (e.g., \citealt{lim2009}). Assuming that realistic observations require $S/N\gtrsim5$ for $\Gamma(u,v)=1$ and observing times $t=1$\,h, we may conclude that the limiting photometric magnitude for an intensity interferometer with the technical parameters of the NSII is $m_X\approx2.5$, $X=V,R,I$.\footnote{\citet{hanbury1974b} actually assumed $S/N\geq3$ at $B$ band, resulting in a limiting magnitude $m_B\approx2.5$.}

\subsubsection{Photon Degeneracy Limit \label{sssect_coherencelimit}}

An additional sensitivity constraint is imposed by the properties of the source of radiation. The \emph{maximum} signal-to-noise ratio is limited by the number of correlated photons actually emitted by the target. Accordingly, it is possible to derive (\citealt{goodman1985}, their Chapter 9.5) the \emph{photon degeneracy limited signal-to-noise ratio}
\begin{equation}
\left(\frac{S}{N}\right)_{\rm p} = \alpha\,\eta\,\Gamma(u,v)\,\left(\frac{t\,\Delta f}{2}\right)^{1/2}\,\chi
\label{eq_snphoton}
\end{equation}
where $\chi$ is the \emph{degeneracy parameter}, i.e., the average number of photons occurring in a single coherence interval of the incident radiation. For the well-known case of blackbody radiation, Bose-Einstein statistics leads to
\begin{equation}
\chi = \left[\exp\left(\frac{h\nu}{kT}\right)-1\right]^{-1}
\label{eq_chi}
\end{equation}
where $h$ is Planck's constant, $k$ is Boltzmann's constant, and $T$ is the thermodynamic temperature of the emitter. Provided that the optical bandpass $\Delta\nu$ is sufficiently small, Equation (\ref{eq_chi}) can also be applied to non-thermal radiation with an \emph{effective} temperature. The effective temperature is defined as the temperature of blackbody radiation with the same intensity, at the given wavelength, as the non-thermal radiation \citep{gamo1966}.

\begin{table}[t!]
\centering
\caption{Photon degeneracy limited source temperatures \label{tab_limits1}}
\begin{tabular}{cccc}
\toprule
Filter & $\lambda$ [nm] & $T^{\rm a}$ [K] & $T^{\rm b}$ [K] \\ \midrule
$U$ & 360 & 5380 & 3420 \\
$B$ & 440 & 4400 & 2800 \\
$V$ & 550 & 3520 & 2240 \\
$R$ & 640 & 3030 & 1930 \\
$I$ & 790 & 2450 & 1560 \\ \bottomrule
\end{tabular}
\tabnote{
Temperatures $T$ are the minimum source temperatures required for achieving $(S/N)_{\rm p}=10$, as function of observing wavelength $\lambda$ (Equations (\ref{eq_snphoton}) and (\ref{eq_chi})). For each wavelength, we quote two limiting temperatures for two sets of parameters.
\\ $^{\rm a}$ For $\Gamma(u,v)=1$, $t=1$\,h, $\alpha=0.2$, $\eta=0.2$, $\Delta f=100$\,MHz.
\\ $^{\rm b}$ For $\Gamma(u,v)=1$, $t=1$\,h, $\alpha=0.8$, $\eta=0.2$, $\Delta f=32$\,GHz.
}
\end{table}

From Equations (\ref{eq_snphoton}) and (\ref{eq_chi}) it is evident that $(S/N)_{\rm p}$ is a strong function of source temperature and observing wavelength. Assuming that a practical astronomical observation requires $(S/N)_{\rm p}\gtrsim10$ at $\Gamma(u,v)=1$ within one hour of observing time, the necessary effective temperatures are on the order of several thousand Kelvin throughout the optical spectrum. In Table~\ref{tab_limits1}, we present the temperatures required for $(S/N)_{\rm p}=10$ for various optical wavelengths. Both sets of temperatures assume $\Gamma(u,v)=1$ and one hour of observing time, but different instrumental parameters -- the first one (case a) corresponding to the NSII, the second one (case b) to an improved instrument. The resulting values provide two conclusions: firstly, even vast technical improvements ease the temperature limits by factors less than two. Secondly, optical intensity interferometry \emph{is limited to hot astronomical targets} -- mostly stars. With limiting temperatures $\gtrsim$2\,000\,K, all stellar spectral types, as well as all stellar luminosity classes, are observable \citep{drilling2000}.

\subsection{Applications \label{ssect_applications}}

The overview provided in the previous sections leaves us with an ambivalent picture of optical intensity interferometry.

On the one hand, intensity interferometry is a very \emph{simple} and \emph{robust} technique. As it processes electronic signals instead of combining light rays directly, intensity interferometers can, essentially, be assembled from technology well-known for radio interferometers, at compatible cost and effort. For realistic electronic bandwidths, coherence lengths are on the order of tens of centimeters; accordingly, intensity interferometers are very tolerant with respect to mechanical aberrations and uncertainties in the coordinates $u,v,$ and $w$. As a side effect, intensity interferometers do not require high-quality telescopes as light collectors but rather simple ``light buckets'', meaning coarse (with tolerances on the order of centimeters) parabolic reflectors, in analogy to atmospheric Cherenkov telescope arrays which have aperture diameters of tens of meters \citep{hanbury1974b,lorenz2004,aharonian2006,lacki2011}. The insensitivity with respect to atmospheric turbulence provides an additional advantage over optical amplitude interferometry. Furthermore, optical intensity interferometers are very \emph{powerful} scientific instruments: they are the \emph{only} optical interferometers that can achieve baseline lengths on scales of kilometers; as intensity interferometers do not need to combine light rays directly, their baselines are not limited in length. Accordingly, the angular resolution $\theta$ of optical intensity interferometers, in practical units, can be written as
\begin{equation}
\frac{\theta}{\mu\rm as} = 176 \left(\frac{\lambda}{700\,\rm nm}\right)\left(\frac{d}{1\,\rm km}\right)^{-1}
\label{eq_theta}
\end{equation}
when using the Rayleigh criterion (cf. Equation (\ref{eq_disk})). A hypothetical global intensity interferometer array with $d\approx10\,000$\,km could achieve angular resolutions on the order of tens of \emph{nano}seconds of arc -- a sufficiently bright and spatially concentrated science target provided.

On the other hand, optical intensity interferometry is severely handicapped by its limitations in sensitivity. The \emph{photon degeneracy limit} restricts the technique to \emph{hot} astronomical targets, especially stars. The \emph{statistical sensitivity limit} restricts intensity interferometry to \emph{bright} astronomical targets. Observatories of the type of the Narrabri interferometer are limited to apparent magnitudes $\lesssim$2.5; accordingly, the NSII was eventually restricted to observations of 32 stars \citep{hanbury1974b}.

In summary, we see that optical intensity interferometry bears the potential of being a very important and powerful astronomical technique \emph{if} its statistical sensitivity (Equations (\ref{eq_snii}) and (\ref{eq_snnsii})) can be improved substantially at acceptable cost and effort. In recent years, multiple studies have noted this possibility and have explored possible applications for modern optical intensity interferometers \citep{lebohec2006,ofir2006a,ofir2006b,klein2007,solomos2008,foellmi2009,barbieri2009,dravins2010a,dravins2010b,lebohec2010,nunez2012a,nunez2012b,dravins2012,dravins2013,rou2013}, including the implementation of a working group within the International Astronomical Union \citep{barbieri2009} and of the dedicated Star Base Utah test facility near Salt Lake City, USA \citep{lebohec2010}. These studies largely exploit the remarkable technical similarities between optical intensity interferometers and Cherenkov air shower telescopes with respect to light collection, photo-detection, and electronic signal processing. Current and future Cherenkov telescopes (are expected to) have collecting areas on the order of several hundred to several thousand square meters \citep{lorenz2004,aharonian2006,nunez2012a,dravins2013}; using these telescopes in a -- secondary -- intensity interferometry mode permits sensitivities higher by factors of several hundred compared to the NSII (which had collecting areas of $\approx$30\,m$^2$; see Equation (\ref{eq_snnsii})). In addition, the sensitivity of astronomical interferometers -- regardless of type -- scales with the number of baselines, $N_{\Join}$, like $(S/N)_{\rm s}\propto\sqrt{N_{\Join}}$ due to improved sampling of the $uv$ plane (cf. Section \ref{ssect_coherence}); accordingly, adding additional light collectors to an array improves the sensitivity further.

\section{Multi-Channel Intensity Interferometry \label{sect_multiii}}

Complementary to the studies mentioned in Section \ref{ssect_applications} which largely concentrate on collecting areas and numbers of baselines, we now focus on the factor
\begin{equation}
\aleph \equiv \left(\frac{\alpha}{0.2}\right)\left(\frac{\Delta f}{100\,{\rm MHz}}\right)^{1/2}
\label{eq_aleph}
\end{equation}
taken from Equation (\ref{eq_snnsii}). The reference values for the quantum efficiency, $\alpha=0.2$, and the electronic bandwidth, $\Delta f = 100$\,MHz, correspond to the case of the NSII \citep{hanbury1974b}. These values originate from the use of photomultiplier tubes and represent the state of the art of the 1970s. Modern, commercially available PMT show values up to $\alpha\approx0.25$ and $\Delta f\approx 1$\,GHz (e.g., \citealt{renker2007}), meaning $\aleph\approx4$ -- a noteworthy, albeit not substantial improvement.

Evidently, we need to apply more sophisticated methods of photo-detection in order to increase $\aleph$ substantially. Especially promising for our purpose are semiconductor, specifically silicon, avalanche photodiodes (Si-APDs -- e.g., \citealt{renker2007,kitchin2009}). With $\alpha\approx85$\%, they outperform PMTs by factors $\approx$4 with respect to quantum efficiency. The quantum efficiency of Si-APDs is a strong function of wavelength and peaks around 700\,nm; accordingly, any astronomical instrument making use of Si-APDs is sensitive mostly in $R$ band.

Even more important than the high quantum efficiency of APDs is their electronic bandwidth. Firstly, APDs provide maximum sampling rates similar to modern PMTs, meaning electronic bandwidths up to $\Delta f\approx1$\,GHz. Secondly, APDs are sufficiently small -- few millimeters -- to be operated in detector \emph{arrays}, with each array pixel corresponding to one APD (e.g., \citealt{anderhub2011}). As already noted by \citet{hanbury1974b}, this can be used to increase the instrumental sensitivity by observing \emph{spectrally dispersed} light from the source with multiple photodetectors simultaneously at different wavelengths -- i.e., in multiple spectral channels.\footnote{This approach requires $\Delta\nu>\Delta f$, a condition which is very difficult to violate at visible wavelengths.}

The basic idea of \emph{multi-channel intensity interferometry} follows from Equation (\ref{eq_snii}) in a straightforward manner. The sensitivity of an intensity interferometer is proportional to the number of photons \emph{per unit bandwidth} and independent of the \emph{total} number of photons. We assume that the light from the source is spectrally dispersed at two light collectors before reaching the photodetectors. For simplicity, we further assume that observations are limited to a sufficiently small part of the source spectrum such that both $n_{\nu}$ and $\alpha$ may be regarded as approximately constant -- usually meaning few tens of nanometers in wavelength, avoiding known spectral lines. At both light collectors ``1'' and ``2'' we divide the spectral band into $N_{\diamond}$ channels of equal width which are monitored by detectors $i=1,2,...,N_{\diamond}$; accordingly, we record intensities $I_{1,i}$ and $I_{2,i}$. We now apply the signal processing scheme outlined in Figure \ref{fig_ii}b to each of the $N_{\diamond}$ spectral channels separately. For each $i$, we compute correlation factors $\Gamma_i(u,v)$ according to Equation (\ref{eq_gamma}). Eventually, we obtain a combined correlation factor from averaging over all spectral channels, meaning $\Gamma(u,v)=\aa\Gamma_i(u,v)\rr_i$. As we average over $N_{\diamond}$ independent measurements, the statistical signal-to-noise ratio of $\Gamma(u,v)$ is higher than that of any $\Gamma_i(u,v)$ by a factor $\sqrt{N_{\diamond}}$. Following \citet{hanbury1974b}, we express this relation in terms of an \emph{effective} electronic bandwidth
\begin{equation}
\Delta f \longrightarrow \Delta f' \equiv N_{\diamond}\,\Delta f
\label{eq_effband}
\end{equation}
which enters into Equations (\ref{eq_snii}), (\ref{eq_snnsii}), (\ref{eq_snphoton}), and (\ref{eq_aleph}) accordingly.

Avalanche photodiodes can be combined into detector arrays similar to early charge-coupled device (CCD) detector arrays; the largest APD arrays in use comprise about 1\,000 pixels (e.g., \citealt{anderhub2011,anderhub2013}). Those APD arrays can be placed inside a spectrometer and record the dispersed source light; each detector pixel along the direction of dispersion then corresponds to one spectral channel. Else than for the case of CCDs, which have electronic bandwidths of few kHz at best, each pixel of an APD array provides a bandwidth on the order of few hundred MHz up to about one GHz. Accordingly, even small APD arrays are able to improve $\Delta f'$, and thus $\aleph$, substantially.

In order to achieve a realistic estimate of $\aleph$, we have to regard the quantum efficiency as well as the effective electronic bandwidth. For Si-APDs, maximum quantum efficiencies are $\alpha\approx85$\% (at $\lambda\approx700$\,nm); this value is set by solid-state physics and, evidently, is already close to optimum. The effective electronic bandwidth is -- a priori -- only limited by the size of APD arrays and the abilities of the signal processing electronics. Signal processing systems of current long baseline radio interferometers are able to handle signal bandwidths up to 32\,GHz (e.g., \citealt{schuster2008,boissier2009}). In case of an optical intensity interferometer equipped with Si-APD arrays, this value may be achieved by using, e.g., 64 spectral channels of 500\,MHz each. Assuming thus, conservatively, $\alpha=0.8$ and $\Delta f'=32$\,GHz (as we did for Table~\ref{tab_limits1}), we find $\aleph=72$. We note that the bandwidth of 32\,GHz for radio interferometers is largely dictated by heterodyne receiver technology and not by the backends. Therefore, a careful extrapolation of the effective electronic bandwidth from 32\,GHz to 64\,GHz (e.g., 128 channels $\times$ 500\,MHz) provides a realistic estimate for the capabilities of a modern optical intensity interferometer, leading to $\aleph=101$ (again, for $\alpha=0.8$). We therefore argue that it is possible to \emph{increase the sensitivity of an NSII type optical intensity interferometer by a factor of approximately 100 by employment of existing photodetector and electronic signal processing technologies.}

An improvement of instrumental sensitivity by a factor of 100 corresponds to five astronomical magnitudes. Accordingly, the limiting magnitude of an NSII-like intensity interferometer improves to $m_R\approx7.5$ (cf. the discussion following Equation (\ref{eq_snnsii})). Due to the well-known inverse-square-of-distance law of radiation flux, such an improved interferometer is able to observe targets ten times further away than the targets of the NSII, accordingly surveying a volume 1\,000 times larger than the survey volume of the NSII. Even when assuming that, as with the NSII, only main sequence and giant stars are observable, the number of potential astronomical targets increases from few tens to \emph{tens of thousands.} Of course, increasing the area of the light collectors and/or adding additional light collectors improves the instrumental sensitivity further.

\section{Generic Instrument Layout \label{sect_instruments}}

The basic design of a practical science-grade optical multi-channel intensity interferometer (MCII) follows from an appropriate combination of techniques from optical and radio observatories. The key optical component is a spectrograph fitted with an APD array for dispersion and recording of the infalling light, respectively. In the following, we assume that such a device is located either in the prime focus or in the Cassegrain focus of a reflecting parabolic light collector (or parabolic antenna in radio astronomical terminology). The key component contributed by radio astronomy is the design of radio interferometric arrays plus signal processing electronics, especially correlators. As already noted in Section \ref{sect_multiii}, we can safely assume that the necessary electronics is available ``off the shelf''; accordingly, we focus on the layout of antennas and arrays. In the following, we are going to provide the figures of merit for realistic MCIIs.

\subsection{Detectors \label{ssect_detectors}}

Commercially available\footnote{Referring here again to the {\sc Hamamatsu Photonics} catalog.} Si-APDs with electronic bandwidths of $\Delta f\approx500$\,MHz as required for an MCII have photo-sensitive areas with diameters of around one millimeter. Individual detectors can be combined into arrays. Arguably the most sophisticated APD array design for astronomical applications is employed by the First APD Cherenkov Telescope (FACT; \citealt{anderhub2011,anderhub2013}) on La Palma (Canary Islands, Spain). The FACT camera uses an array composed of 1440 individual APDs -- a number one order of magnitude larger than the one required for an MCII; accordingly, FACT has already demonstrated the feasibility of large APD detector arrays in astronomical instruments. In case of an MCII camera, APDs are to be arranged in a linear array along the direction of dispersion. Assuming an array with 128 detectors of 1\,mm diameter each, and permitting for an additional margin of 50\% for gaps between detectors, implies detector arrays spanning up to approximately 20\,cm in length.

In order to prevent a serious degradation of the overall instrument efficiency (to be discussed in Section~\ref{ssect_eff}), it is probably necessary to cool the APDs to temperatures below $-30\deg$C to limit their dark count rate to about 10\,000 electrons per second and per square millimeter of detector surface.\footnote{Assuming here a dark count rate of $10^6$ electrons~s$^{-1}$\,mm$^{-2}$ at $+30\deg$C for high-quality APDs \citep{biland2014} and a decrease of the count rate by a factor 1.08 for each degree Kelvin the temperature is reduced (\textsc{Hamamatsu} documentation).}

\subsection{Spectrographs \label{ssect_spectrographs}}

The \emph{spectral resolution} required for an MCII is governed by the width $\Delta\lambda$ of the spectral band to be examined and by the number of spectral channels. Assuming observations at a wavelength $\lambda=700$\,nm through a narrow-band filter with a bandpass of $\Delta\lambda=20$\,nm implies a spectral resolution of $\lambda/\Delta\lambda=35$ by the narrow-band filter alone. This number needs to be multiplied by the number of spectral channels. Assuming an APD array with 128 detectors leads to a required spectral resolution of $\lambda/\Delta\lambda\geq4500$; this is a value readily provided by simple spectrometers.

A much stronger constraint is imposed by the necessary (reciprocal) \emph{linear dispersion} ${\rm d}\lambda/{\rm d}s$, with $s$ denoting the physical distance along the spectrum in the detector plane (using the definitions provided by Equation (4.28) of \citealt{kitchin2009}). Dispersing a band of 20\,nm along a detector spanning 20\,cm in size corresponds to a linear dispersion of 0.1\,nm\,mm$^{-1}$; this is actually the lower end of the interval covered by practical astronomical spectrometers. Accordingly, MCII spectrometer designs need to balance (i) the width of the spectral band $\Delta\lambda$, (ii) the number of spectral channels, (iii) the size of individual APDs, and (iv) the overall size and complexity of the instrument.

\subsection{Light Collectors \label{ssect_collectors}}

To first order, a practical MCII light collector can be designed like a parabolic radio antenna with a surface cover -- like aluminum or silver -- that reflects visible light efficiently. The mechanical tolerance limits to be obeyed follow from (i) the electronic bandwidth $\Delta f$ and (ii) the instrumental point spread function (PSF).

According to Equation (\ref{eq_taucii}), the coherence length of an intensity interferometer with electronic bandwidth $\Delta f$ is $w'\approx c/\Delta f$. In order to prevent a substantial reduction of the degree of correlation, the error on the total optical path length must not exceed $\approx$10\% of this value (cf. Figure \ref{fig_g2}); for $\Delta f=500$\,MHz, this implies an overall tolerance of about 6\,cm including, especially, the maximum deviation of the antenna surface from a mathematical paraboloid.\footnote{This excludes the use of Davies--Cotton light collectors \citep{davies1957} that are employed by some Cherenkov telescopes (e.g., HESS; \citealt{aharonian2006}).}

The light collected by the reflector needs to be focused onto the entrance pupil (or slit) of the spectrometer. In order to ensure that neighboring APDs in the detector array are indeed independent spectral channels, monochromatic images of the entrance pupil must not exceed individual APDs in size. Calculating conservatively, this limits the diameter of the entrance pupil, and thus the size of the image of the instrumental point spread function in the focal plane, to about one millimeter. By geometry, pupil diameter $p$, focal length $F$, and angular diameter $\beta$ of the PSF (which equals the field of view of the pupil in our calculation) are related like $\beta=p/F$ (in small--angle approximation). For $p=1$\,mm, this implies values for $\beta$ of $206''$, $41''$, $21''$, and $10''$ for focal lengths of 1\,m, 5\,m, 10\,m, and 20\,m, respectively; these limits might be relaxed by factors up to about three by use of dedicated focal reducers (e.g., \citealt{lim2013}). For telescopes with focal ratios around unity (as common in radio and TeV/Cherenkov astronomy) this implies that 10-meter class light collectors need to limit the size of their instrumental PSFs, as well as pointing/tracking uncertainties, to tens of arcseconds -- which is achieved regularly by modern radio telescopes (e.g., \citealt{wilson2010}).

\subsection{Interferometer Arrays \label{ssect_arrays}}

The design of a science-grade MCII array is mainly governed by two boundary conditions. Firstly, the observatory should provide for a sufficient signal-to-noise ratio (referring here to the statistical limit as given by Equations (\ref{eq_snii}) and (\ref{eq_snnsii})). Secondly, the observatory should comprise multiple baselines in order to permit for an analysis of the, potentially complicated, 2-dimensional structure of a given target source.

The sensitivity (statistical signal-to-noise ratio) of an array of $M$ antennas follows from Equations (\ref{eq_snii}) and (\ref{eq_snnsii}) via
\begin{equation}
\label{eq_snarray}
\left(\frac{S}{N}\right)_{\rm array} = N_{\Join}^{1/2}\left(\frac{S}{N}\right)_{\rm s}
\end{equation}
where $N_{\Join} = M(M-1)/2$ is the number of baselines (Equation (\ref{eq_snarray}) is equivalent to the \emph{radiometric formula} for radio interferometers; e.g., Equation (6.62) of \citealt{wilson2010}). We note that we always use the \emph{effective} electronic bandwidth for calculating signal-to-noise ratios via Equation (\ref{eq_snarray}) (cf. Equation (\ref{eq_effband})). In order to provide realistic examples, we explore four, increasingly complex, interferometer designs in the following; all limiting $R$-band magnitudes quoted below assume $S/N=5$, observing time $t=1$\,h, $\alpha=0.8$, $\eta=0.2$, and $\Gamma(u,v)=1$. The different designs are summarized in Table~\ref{tab_arrays}.

\begin{table}[t!]
\centering
\caption{Parameters of practical optical MCII arrays \label{tab_arrays}}
\begin{tabular}{lrrrr}
\toprule
Layout & $M$ & $D$ [m] & $\Delta f'$ [MHz] & $m_R$ \\
\midrule
portable &  3 &  2 &   8$\times$500 &  4.0 \\
miniII   &  3 & 10 &  32$\times$500 &  8.3 \\
oNOEMA   & 12 & 15 & 128$\times$500 & 11.6 \\
oVLA     & 27 & 25 & 128$\times$500 & 13.6 \\
\bottomrule
\end{tabular}
\tabnote{
{\sc Parameters:} $M$: number of light collectors; $D$: light collector diameter; $\Delta f'$: effective electronic band width (number of channels $\times$ bandwidth per channel); $m_R$: limiting $R$--band magnitude (assuming $S/N=5$, $\Gamma(u,v)=1$, $\alpha=0.8$, $\eta=0.2$, $t=1$\,h).
}
\end{table}

\subsubsection{Portable \label{sssect_portable}}

The insensitivity of intensity interferometers with respect to mechanical aberrations permits for portable MCIIs that can be transported to different locations (``portable'' here means that the entire observatory can, in principle, be stored in a standard $2.4\times2.4\times12$\,m$^3$ freight container). Such an array should comprise three light collectors -- the minimum number needed to span a plane, providing three baselines -- with diameters $D=2$\,m (collecting area $A=3$\,m$^2$) each. Improved $uv$ coverage may be achieved by repeated observations of the same target with different arrangements of the light collectors. In order to keep the spectrometer reasonably small and simple, we assume the use of eight 500-MHz channels, corresponding to an effective electronic bandwidth of 4\,GHz. From Equation (\ref{eq_snarray}) we find a limiting $R$-band magnitude $m_R=4.0$. We note that already this simple interferometer layout comes with a sensitivity limit better than the one of the Narrabri Stellar Intensity Interferometer by more than one photometric magnitude.

\subsubsection{miniII \label{sssect_miniii}}

A minimalist intensity interferometer (miniII) layed out for a substantial science program should employ three light collectors (providing three baselines) with diameters of about ten meters (collecting areas of about 75\,m$^2$) each. The light collectors could be made movable by placing them on tracks, thus permitting for improved $uv$ coverages by repeated observation of the same target with different array configurations. In order to balance instrumental sensitivity and complexity, we assume the use of 32 channels of 500\,MHz electronic bandwidth each, i.e., an effective electronic bandwidth of 16\,GHz. Collecting area, number of baselines, and electronic bandwidth imply (via Equation (\ref{eq_snarray})) a limiting magnitude $m_R=8.3$.

\subsubsection{oNOEMA \label{sssect_onoema}}

Modern long-baseline radio interferometers provide good templates for designs of MCII arrays. An example for an array of intermediate size is provided by the Northern Extended Millimeter Array (NOEMA; \citealt{schuster2008}) currently under construction on the Plateau de Bure (Hautes-Alpes, France).\footnote{See \tt http://www.iram-institute.org/EN/content-page- 261-9-261-0-0-0.html \rm for the observatory status.} A corresponding ``optical NOEMA'' (oNOEMA) would comprise 12 light collectors of 15 meters diameter each, meaning a collecting area of 170\,m$^2$ per collector and 66 baselines in total. We assume a sophisticated detector system with 128 spectral channels of 500\,MHz bandwidth each, providing an effective electronic bandwidth of 64\,GHz. Accordingly, the limiting magnitude is $m_R=11.6$.

\subsubsection{oVLA \label{sssect_ovla}}

The Karl G. Jansky Very Large Array\footnote{See \tt http://www.vla.nrao.edu/ \rm for details.} (VLA) near Socorro (New Mexico, USA) is one of the largest long-baseline radio interferometers in operation, equipped with 27 antennas and spanning 36\,km in size. A corresponding ``optical VLA'' (oVLA) would comprise 27 light collectors of 25\,m diameter each, meaning a collecting area of 470\,m$^2$ per collector and 351 baselines in total. When assuming 128 spectral channels of 500\,MHz bandwidth each (i.e., an effective electronic bandwidth of 64\,GHz), one finds a limiting magnitude $m_R=13.6$.

\subsubsection{oVLBI \label{sssect_ovlbi}}

The interferometer array layouts described in the previous paragraphs implicitly assume long-baseline interferometry with all light collectors being placed at the same observatory site. As noted in Section \ref{ssect_radio}, existing very long baseline array (VLBI) techniques overcome this limitation. State-of-the-art VLBI recorders are able to store data at a speed corresponding to an electronic bandwidth of 4\,GHz \citep{whitney2013}; accordingly, this value is a practical limit for the electronic bandwidth (of a single spectral channel) of an MCII. Optical VLBI (``oVLBI'') arrays could achieve larger \emph{effective} electronic bandwidths by using more than one recorder per light collector, with each recorder storing the output of one or more spectral channels.

\subsection{Instrument Efficiencies \label{ssect_eff}}

Throughout this paper we assume a generic instrument efficiency (excluding the quantum efficiency $\alpha$ of the detectors) $\eta=0.2$ which we can motivate as follows. We approximate the instrument efficiency as the product of the efficiencies resulting from three dominating effects. Firstly, the optical efficiency of astronomical spectrometers is about $\eta_{\rm spec}\approx0.7$ (\citealt{kitchin2009}, their Chapter 4.2.1). Secondly, the fraction of the light that is actually collimated onto the spectrometer pupil and makes it into the spectrometer, i.e., the collimation efficiency (or aperture efficiency) of the light collector, can be estimated from modern radio (e.g., \citealt{kim2011,lee2014}) and Cherenkov telescopes (e.g., \citealt{aharonian2006}) and is located in the range $\eta_{\rm coll}\approx0.5-0.7$. Thirdly, the dark count rate of the APDs, which contributes to the total count rate. Assuming a photon flux of $n_{\nu}=9\times10^{-5}\,{\rm m}^{-2}\,{\rm s}^{-1}\,{\rm Hz}^{-1}$ for an $m_R=0$ star (cf. Section~\ref{sssect_statlimit}), a bandpass of $\Delta\lambda=0.1$\,nm for any given APD (cf. Section~\ref{ssect_spectrographs}), a dark count rate of 10\,000 electrons per seconds for each 1-mm$^2$ sized APD (cf. Section~\ref{ssect_detectors}) corresponds to $\approx$3\% (for a portable MCII) to $\approx$100\% (for the oVLA design) of the photon rate from a source at the limiting magnitude given in Table~\ref{tab_arrays}. This translates into an efficiency factor contributed by the dark count rate varying from $\eta_{\rm dark}\approx0.97$ to $\eta_{\rm dark}\approx0.5$. In summary, we may expect instrument efficiencies $\eta=\eta_{\rm spec}\,\eta_{\rm coll}\,\eta_{\rm dark}\approx0.2-0.5$.

\subsection{Observatory Sites \label{ssect_site}}

Compared to the needs of optical observatories, the conditions to be fulfilled by the site for an MCII are less demanding in general. As already pointed out in Section \ref{ssect_coherencetimes}, intensity interferometers are insensitive to atmospheric fluctuations, meaning that atmospheric seeing is essentially irrelevant for choosing a location. The most important limitation for an MCII is the night sky brightness. High-quality astronomy sites (referring here specifically to Cerro Paranal, Chile; \citealt{patat2004}) show a sky brightness of approximately 12 magnitudes per square arcminute in $R$-band. As noted in Section \ref{ssect_collectors}, practical MCIIs may be expected to have effective fields of view between few hundred and few thousand square arcseconds; accordingly, even at a very dark site, the sky contributes light, and thus photon noise, corresponding to a source with $12\lesssim m_R\lesssim15$. For MCIIs aiming at sources with $m_R\gtrsim12$, like the oNOEMA or oVLA layouts of Section \ref{ssect_arrays}, this is about the same as the photon count rate due to the actual science target. Accordingly, the night sky brightness might effectively double the noise and thus decrease the sensitivity by a factor on the order of two, corresponding to 0.8 photometric magnitudes. Consequently, MCIIs aimed at faint targets require a careful site selection, possibly accompanied by an optical design which minimizes the field of view of the spectrometer pupil.

\section{Science Cases \label{sect_science}}

\subsection{Stellar Diameters \label{ssect_diameters}}

The direct measurement of the angular diameters of stars has been the most important driver for optical interferometry since the pioneering work by \citet{michelson1921}, and especially, of course, for the development of optical intensity interferometry \citep{hanbury1964,hanbury1967b}. Precise measurements of stellar angular diameters\footnote{Those measurements actually require distinction between the \emph{uniform disk diameter} $\theta_{\rm UD}$, which follows from modeling the stellar intensity distribution as a uniformly illuminated disk, and the \emph{limb-darkened diameter} $\theta_{\rm LD}$ which includes the effects of limb-darkening. For main-sequence stars, $\theta_{LD}$ is larger than $\theta_{\rm UD}$ by about 2--4\% (e.g., \citealt{boyajian2012a,boyajian2012b}).} are crucial for accurate (to better than few per cent) determinations of linear radii, effective temperatures, and luminosities of stars (cf. \citealt{boyajian2012a,boyajian2012b,boyajian2013}). Even though angular diameters have been derived for over 8\,000 stars to date \citep{richichi2005,boyajian2012a}, only few hundred measurements with accuracies better than 5\% are available. In addition, only about 10\% of all stars probed are main-sequence stars covering the spectral classes A to M; only a handful of O and B type main-sequence stars have been measured (by the Narrabri intensity interferometer; \citealt{hanbury1974b}).

As already hinted at in Section \ref{sect_multiii}, multi-channel intensity interferometers with limiting magnitudes $m_R\gtrsim8$ have tens of thousands of potential targets. Evidently, the distance up to which a star can be observed is a function of limiting apparent magnitude $m_R$ as well as absolute magnitude $M_R$; accordingly, the range $r$ of a given intensity interferometer is a function of stellar type. In addition, the angular diameter $\theta$ of a star at a given distance $r$ depends on its linear radius $R_{\star}$; accordingly, the required angular resolution of the interferometer -- here assumed to be given by the Rayleigh criterion, $\theta=1.2\lambda/b$, with $b$ being the maximum baseline length, and $\lambda=700$\,nm -- likewise depends on the stellar type.

In Table~\ref{tab_ranges}, we summarize the ranges and angular resolutions of interferometers aimed at main-sequence stars of types O to M, assuming two different limiting magnitudes $m_R=8$ and $m_R=12$, respectively. We note that the actual values for $r$ can be reduced substantially by interstellar extinction (which in turn can be compensated partially by longer observing times). The baseline lengths required for fully resolving a given star are on the order of kilometers typically, in agreement with the sizes of modern long-baseline radio interferometers. Observational ranges vary from the immediate solar neighborhood (tens of parsecs) for M stars to substantial fractions of the Milky Way (thousands of parsecs) for O and B stars; indeed, sensitive ($m_R\gtrsim12$) interferometers are able to analyze O stars throughout the Milky Way. Simple upscaling of the values provided in Table~\ref{tab_ranges} tells us that interferometers with a limiting magnitude $m_R>13$ and baselines on the order of tens of kilometers, along the lines of the oVLA concept (cf. Table~\ref{tab_arrays}), are able to resolve O-type main sequence stars in the Magellanic Clouds.\footnote{This is aided by low foreground extinction toward the Magellanic Clouds, with $A_V\approx0.2$\,mag \citep{larsen2000}.}

\begin{table}[t!]
\centering
\caption{Interferometer observation ranges for main sequence stars \label{tab_ranges}}
\begin{tabular}{lrrrrrr}
\toprule
Type & $M_R$ & $R_{\star}$~  & $m_R$ & $r$~  & $\theta$~~ & $b$~~ \\
     &       & $[R_{\odot}]$ &       & [pc]  & [$\mu$as]  & [km]  \\
\midrule
O5V & $-$5.5 & 12  &  8 &  5\,000 &  22 &  7.8 \\
    &        &     & 12 & 31\,600 &   4 & 49.0 \\ \addlinespace
B0V & $-$3.9 & 7.4 &  8 &  2\,400 &  29 &  6.0 \\
    &        &     & 12 & 15\,000 &   5 & 38.0 \\ \addlinespace
A0V &    0.6 & 2.4 &  8 &     300 &  74 &  2.4 \\
    &        &     & 12 &  1\,900 &  12 & 15.0 \\ \addlinespace
F0V &    2.4 & 1.5 &  8 &     130 & 110 &  1.6 \\
    &        &     & 12 &     830 &  17 & 10.0 \\ \addlinespace
G2V &    4.2 &  1  &  8 &      58 & 160 &  1.1 \\
    &        &     & 12 &     360 &  26 &  6.7 \\ \addlinespace
K0V &    5.3 & 0.9 &  8 &      35 & 240 &  0.7 \\
    &        &     & 12 &     220 &  38 &  4.5 \\ \addlinespace
M2V &    8.4 & 0.5 &  8 &       8 & 580 &  0.3 \\
    &        &     & 12 &      52 &  89 &  1.9 \\
\bottomrule
\end{tabular}
\tabnote{
{\sc Parameters:} ``Type'': spectral type of target star; $M_R$: absolute $R$-band magnitude; $m_R$: maximum apparent $R$-band magnitude; $R_{\star}$: stellar radius in units of solar radius $R_{\odot}$; $r$: maximum distance to target star; $\theta$: angular diameter of the star at distance $r$; $b=1.2\lambda/\theta$, for $\lambda=700$\,nm. Star data are from \citet{drilling2000}.
}
\end{table}

\subsection{White Dwarfs \label{ssect_dwarfs}}

Direct measurements of the radii of white dwarfs provide independent tests of theoretical mass--radius relationships of compact stellar remnants (see, e.g., \citealt{holberg1998,barstow2005} for the case of Sirius B). A census of the solar neighborhood\footnote{Referring here to the RECONS data base, {\tt www.recons.org}} finds six white dwarfs with $m_V<12.5$ within 5.5\,pc from the sun, the nearest and brightest one being Sirius B with $m_V=8.4$ located at a distance of 2.6\,pc.

At a distance of 5.5\,pc, a white dwarf with a diameter of 10\,000\,km has an angular diameter of 12\,$\mu$as. This corresponds to the angular resolution of an MCII array with a maximum baseline length of 14 kilometers (for $\lambda=700$\,nm).

\subsection{Stellar Sub-Structure \label{ssect_surfaces}}

In general, the measurement of stellar diameters (Section \ref{ssect_diameters}) assumes simple distributions of the stellar light on sky, usually uniform circular disks with limb darkening. If (i) sufficient two-dimensional $uv$ coverages and (ii) sufficient sensitivities are provided, MCIIs can be used to probe deviations -- like rotation flattening or starspots -- from those simple first-order models. The required sensitivities can be estimated from Equations (\ref{eq_snii}) or (\ref{eq_snnsii}) by replacing the correlation factor $\Gamma(u,v)$ with the difference in correlation relative to a reference model (e.g., the uniform disk model illustrated in Figure \ref{fig_disk}), $\Delta\Gamma(u,v)<1$. When probing structure causing small deviations with $\Delta\Gamma(u,v)=0.01$, the required brightness of the target star increases by a factor 100, corresponding to five photometric magnitudes, compared to simple size measurements. Naturally, details strongly depend on the structure probed, as well as the $uv$ coverage achieved by a given interferometer; a detailed, quantitative description of an analysis of stellar surface structure by means of intensity interferometry is provided by \citet{nunez2012b}. Several specific science cases are discussed in the following Sections \ref{ssect_rotation}, \ref{ssect_convection}, and \ref{ssect_multistars}.

\subsection{Stellar Rotation \label{ssect_rotation}}

Rapid stellar rotation (with rotation speeds in excess of 100\,km\,s${^{-1}}$) leads to (i) apparent elongation of a star due to rotational flattening and (ii) surface gravity darkening, caused by changes of the surface temperature by up to 1\,000\,K. Theoretically, rotation is supposed to play a key role for the structure and evolution of stars, but details are still poorly understood. Systematic studies of surface gravity darkening are important for accurate calibrations of various stellar parameters (cf., e.g., \citealt{zhao2009,deupree2012}). Both the shape (elongation) and the gravity darkening of stellar surfaces may depend on the angular velocity profiles inside stars. Accordingly, interferometric mapping of stellar surfaces could unveil the physical processes responsible responsible for the transport of angular momentum and chemical species inside stars. Specifically, existing observational evidence for differential rotation of fast rotating stars (see, e.g., \citealt{monnier2007} for the case of Altair) is in conflict with dynamical models predicting rigid-body rotation \citep{spruit2002,eggenberger2005,heger2005} -- thus making obvious the need for further and more detailed interferometric studies.

\subsection{Starspots \label{ssect_convection}}

Starspots arise from interactions of stellar photospheres with magnetic fields. In case of cool stars with convective outer layers (roughly, spectral types F--M), the emergence of magnetic field elements in the photosphere reduces convection locally, thus reducing the local temperature and causing the appearance of \emph{dark} starspots. Probably all stars with convective outer layers have dark starspots; observationally, those spots are found to be up to about 20\% cooler than the photospheres and to cover up to approximately 20\% of a stellar surface \citep{strassmeier2009}. Due to the limited angular resolutions of optical amplitude interferometers, interferometric studies of dark starspots are sparse and have been limited to red supergiant stars so far \citep{haubois2009,baron2014}.

Recent studies \citep{cantiello2009,cantiello2011} indicate that sub-surface convection plays a key role also for the surface activity of hot stars with radiative outer layers (roughly, spectral types O--A). Here, magnetically driven bubbles ascending from the convection zones increase the transparency of the photosphere, causing magnetic \emph{bright} starspots. Such bright starspots might be the cause for the observed flux variations of about 0.1\% on timescales of days in O stars \citep{cantiello2011}. Accordingly, interferometric identification of such hot spots would provide a clue on the nature of photometric variability, wind clumping, and X-ray variability of massive stars.

\subsection{Multiple Star Systems \label{ssect_multistars}}

Multiple star systems are a classic target of interferometric observations (see, e.g., \citealt{mcalister1985} for a review), resulting in the discovery of \emph{interferometric binaries} \citep{hanbury1974b} and shedding light on the dynamics of, and the interactions between, the member stars (cf., e.g., \citealt{baron2012} for the case of the Algol system). Dedicated MCII surveys could unveil binary systems consisting of relatively low-mass helium stars of about 2--6 solar masses and OB type stars which are considered typical progenitors of Type Ib/c supernovae \citep{podsiadlowski1992,yoon2010} but have not been observed yet. Mass-accretion onto the mass gainer of an interacting binary system during the mass exchange phase is one of the most important, but poorly understood, physical processes that occur during the evolution of binary stars. Interferometric mapping of the accretion disk in an mass-exchanging binary system may therefore give useful insight on the accretion physics in binary systems: mass-accretion efficiency (i.e., the ratio of the transferred matter to the accreted matter), tidal interactions between the disk and the orbit and the interactions between the transferred matter and the stellar winds from a massive stellar component, et cetera (cf., e.g., \citealt{zhao2008}). This may go as far as to direct observations of the ``accretion belts'' around accreting white dwarfs in binary systems predicted by \citet{kippenhahn1978}.

\subsection{Interstellar Distance Measurements \label{ssect_distances}}

Multi-channel intensity interferometry permits measurements of interstellar distances via several paths. Firstly, and evidently, the distance $r$ to a star with linear radius $R_{\star}$ follows from the angular diameter $\theta$ (Section \ref{ssect_diameters}) via the relation $r = 2R_{\star}/\theta$; accordingly, this approach can be applied to all stars for which the linear radius is known better than the distance -- indeed, sufficiently sensitive MCIIs (cf. Table \ref{tab_ranges}) are, in principle, able to map parts of the Milky Way with accuracies (of about 5\%) currently reserved for radio interferometric astrometry (e.g., \citealt{honma2012}). Secondly, the distance to expanding or pulsating stars can be obtained via the \emph{interferometric Baade--Wesselink method} (e.g., Chapter 3.5.1 of \citealt{degrijs2011}) which relates the (spectroscopically measured) radial velocity to (interferometrically measured) modulations of the angular size. Example science targets are Cepheid stars and, possibly, bright nearby supernovae.

\subsection{Active Galactic Nuclei \label{ssect_agn}}

Active galactic nuclei (AGN) are luminous (up to $10^{15}$ solar luminosities) emitters of light powered by accretion of gas onto supermassive ($10^{7}\lesssim M\lesssim10^{10}\,M_{\odot}$) black holes (e.g., \citealt{beckmann2012}). Quasars and BL Lacertae objects are AGN for which the relative orientation of source and observer permits a direct view along luminous jets into the central engines. The central accretion zone around a black hole crudely spans a few hundred Schwarzschild radii (e.g., \citealt{narayan2005}), corresponding to a few hundred astronomical units or about a thousandth of a parsec for a $10^8$ solar mass black hole; accordingly, these regions have eluded direct observations so far. Interferometric mapping of the central engines of AGN would provide new insights into the physics of black hole accretion, specifically the structure and formation of, and interaction between, accretion disks and relativistic jets, as well as the rotation of massive black holes (e.g., \citealt{mckinney2013}).

Systematic AGN surveys have found a handful of quasars and BL Lac objects with magnitudes $m_V<14$ at any given time (e.g., \citealt{hewitt1993}) which are, in principle, observable with sensitive MCII arrays. The brightest of these, with $m_V=12.8$ in 2009, is the quasar 3C 273 located\footnote{Source data from the NED: {\tt http://ned.ipac.caltech.edu}} at a redshift $z=0.16$. This redshift translates into an image scale of 3\,pc\,mas$^{-1}$; accordingly, a region with a diameter of one milliparsec, corresponding to $\theta=0.3\,\mu$as, is resolved by a very-long baseline interferometer with a maximum baseline length of about 550\,km (for $\lambda=700$\,nm; cf. Equation (\ref{eq_theta})).

We note that the number of potential target AGN is a function of time. Quasars and BL Lac objects show strong, non-periodic flux variability obeying red-noise statistics (cf., e.g., \citealt{park2012,park2014,kim2013} for detailed technical discussions) resulting in brightness variations by up to three photometric magnitudes within about two years (cf., e.g., \citealt{schramm1993} for $R$-band photometry of 3C 345).

\subsection{Amplitude Interferometry \label{ssect_crosscalib}}

Intensity interferometry complements, rather than replaces, amplitude interferometry: intensity interferometry is suited best for observations of bright, hot, and small objects that require long baselines, whereas amplitude interferometry is most useful for observations of large objects where short ($<$1\,km) baselines suffice. Recent results demonstrate the power of amplitude interferometry for constraining the physics of stars and imaging of stellar systems (e.g., \citealt{baron2012,boyajian2012a,boyajian2012b,boyajian2013}) and resolving the structure of nearby AGN (e.g., \citealt{raban2009}).

Amplitude interferometry is, essentially, based on the measurement of fringe visibilities as function of $uv$ coordinates, $V(u,v)\in[0,1]$. Those measurements are highly sensitive to mechanical instabilities of the instrument as well as atmospheric turbulence occurring on timescales of milliseconds (cf. Section \ref{ssect_optical}). Accordingly, visibility measurements require estimates of the \emph{system visibility}
\begin{equation}
\label{eq_visi-sys}
V_{\rm sys} = \frac{V_{\rm cal}^{\rm obs}}{V^{\rm theo}_{\rm cal}}
\end{equation}
where $V_{\rm cal}^{\rm obs}$ and $V^{\rm theo}_{\rm cal}$ denote the observed and theoretical visibility of a calibrator star, respectively. As soon as the system visibility is known, it can be used to derive the \emph{physical} visibility of the science target like
\begin{equation}
\label{eq_visi-phys}
V^{\rm phys}_{\rm target} = \frac{V_{\rm target}^{\rm obs}}{V_{\rm sys}}
\end{equation}
where $V_{\rm target}^{\rm obs}$ denotes the observed target visibility before calibration \citep{vanbelle2005}. 

Remarkably, this calibration scheme implies that interferometric measurements require observations of calibration stars with angular diameters known in advance. The calibrator star diameters have to be obtained from observations of other observatories and/or from theoretical models. Even more remarkably, it appears that, under certain circumstances, those calibration schemes bear the potential for circular conclusions. A recent example is provided by \citet{boyajian2012a} who obtained diameters of main sequence stars with the Mt. Wilson CHARA array. Firstly, they estimated the angular size of their calibration stars by fitting theoretical spectral energy distributions to photometric data. Secondly, they used these calibrator diameters to calibrate the visibility data for the target stars. Thirdly, they compared the observed target star diameters with those expected from theoretical spectral energy distributions.

Intensity interferometers are highly insensitive with respect to mechanical instabilities or atmospheric turbulence. A dedicated calibration of the normalized correlation factors $\Gamma(u,v)$ is not required. Accordingly, it is possible to derive diameters and geometries of a large sample of stars in an unbiased and model-independent way. From this sample, calibrators for amplitude interferometry can be drawn whose diameters are known to be free from the systematic errors that affect the currently applied calibration schemes.

\section{Summary and Conclusions \label{sect_summary}}

\begin{quote}
\small
At the present time this proposed instrument is no more than some drawings and a model. Nevertheless, the scientific rewards of carrying on the work of the Narrabri stellar interferometer would be great; one can only hope that in due course the opportunity to do so will arise.

\hfill --- \citet{hanbury1974b}, p. 172
\end{quote}
We summarized and discussed the theoretical background and concepts of optical Hanbury Brown--Twiss intensity interferometry. This technique is based on the correlation of electronic signals output by photodetectors recording the light received by two (or more) light collectors. Intensity interferometry probes the square modulus of the first-order coherence function of the source light, thus mapping the spatial structure of the target. Due to fundamental quantum limits, intensity interferometry is restricted to observations of hot targets with blackbody temperatures in excess of about 2\,000\,K. We arrive at the following principal conclusions:

\begin{enumerate}

\item  A priori, optical intensity interferometry is a very powerful astronomical technique. Being highly insensitive with respect to mechanical distortions, imperfections of light collectors, and atmospheric turbulence, it is much easier and more economic to operate than optical amplitude interferometry. As intensity interferometers correlate electronic signals (instead of combining light directly as in amplitude interferometers), arbitrary baseline lengths -- up to intercontinental distances -- are possible. Accordingly, it is possible to build and operate large \emph{optical} interferometers at the cost of \emph{radio} interferometers.

\item  Optical intensity interferometry is severely limited by its low sensitivity. The sensitivity can be improved substantially by using arrays of avalanche photodiodes, instead of the historically used single photomultiplier tubes, for light detection. When spectrally dispersing the source light and observing this spectrum with an array of detectors, the sensitivity increases proportional to the square root of the number of independent spectral channels, thus suggesting \emph{multi-channel intensity interferometry}. Additionally, APDs outperform PMTs by a factor of $\approx$4 in terms of quantum efficiency, improving the sensitivity accordingly. As the sensitivity of APDs peaks around $\lambda=700$\,nm, MCIIs are sensitive mostly in $R$-band.

\item  Using conservative estimates based on currently available technology, the sensitivity of multi-channel intensity interferometers with multiple large light collectors -- very similar to modern radio interferometer arrays -- increases by factors up to approximately 25\,000, corresponding to 11 photometric magnitudes, compared to the pioneering Narrabri Stellar Intensity Interferometer. This implies limiting $R$-band magnitudes up to $m_R\approx14$, which is sufficient to observe and resolve main sequence O-type stars located in the Magellanic Clouds. 

\item Sensitive multi-channel intensity interferometry is able to address (i) linear radii, effective temperatures, and luminosities of stars, via direct measurements of stellar angular sizes; (ii) mass--radius relationships of compact stellar remnants, via direct measurements of the angular sizes of white dwarfs; (iii) stellar rotation, via observations of rotation flattening and surface gravity darkening; (iv) stellar convection and the interaction of stellar photospheres and magnetic fields, via observations of dark and bright starspots; (v) the structure and evolution of multiple stars, via mapping of the companion stars and of accretion flows in interacting binaries; (vi) direct measurements of interstellar distances, derived from angular diameters of stars or via the interferometric Baade--Wesselink method; (vii) the physics of gas accretion onto supermassive black holes, via resolved observations of the central engines of luminous active galactic nuclei; and (viii) calibration of \emph{amplitude} interferometers by providing a sample of calibrator stars.

\end{enumerate}  

If implemented eventually, multi-channel intensity interferometry will open a new window for observational astronomy -- and provide an ``easier way to the stars''.


\acknowledgments
We are grateful to \name{Pavel Kroupa} (U Bonn), \name{Sang Gak Lee} (SNU), \name{Philipp Podsiadlowski} (U Cambridge, UK), and \name{Seung Yeon Rhee} (Yale U) for fruitful discussions. This work made use of the software package DPUSER developed and maintained by \name{Thomas Ott} at MPE Garching,\footnote{\tt http://www.mpe.mpg.de/$\sim$ott/dpuser/index.html} the RECONS data base, and the NASA/IPAC Extragalactic Database (NED). We acknowledge financial support from the Korean National Research Foundation (NRF) via Basic Research Grant 2012-R1A1A2041387. SCY was supported by the Research Settlement Fund for new SNU faculty. Last but not least, we are grateful to an anonymous referee for helpful suggestions.



\end{document}